%% file: report.tex
\documentclass[conference]{IEEEtran}
\usepackage{wrapfig}
\usepackage{ifpdf}
\usepackage{balance}
\usepackage{graphicx}
\graphicspath{{./figure/}}
\usepackage{hyphenat}
\usepackage{subfig}
\usepackage{hyperref} 
\usepackage[linesnumbered,ruled,vlined]{algorithm2e}
\SetKwProg{Fn}{Function}{}{end}\SetKwFunction{FRecurs}{Fun}%
\SetAlgoLongEnd

\usepackage{etoolbox}
\usepackage{adjustbox}
\usepackage{amsmath}
\usepackage{amssymb}

\usepackage[dvipsnames]{xcolor}
\definecolor{verde}{rgb}{0.25,0.5,0.35}
\definecolor{jpurple}{rgb}{0.5,0,0.35}
\definecolor{darkgreen}{rgb}{0.0, 0.2, 0.13}
\usepackage{listings}

\definecolor{almond}{rgb}{0.98, 0.98, 0.82}
\definecolor{blue}{rgb}{0.0, 0.0, 1.0}

\newenvironment{myitemize}
{ \begin{itemize}	
		\vspace{-1ex}	
		\setlength{\itemsep}{0pt}
		\setlength{\parskip}{0pt}
		\setlength{\parsep}{0pt}    }
	{ 	 \end{itemize}                    }

\newtheorem{definition}{\textbf{Definition}}

\usepackage{soul} 

\newcommand{\Csharp}{%
	{\settoheight{\dimen0}{C}\kern-.05em \resizebox{!}{\dimen0}{\raisebox{\depth}{\#}}}}
\newcommand{\tony}[1]{{\textcolor{black}{ #1}}}

\newcommand{\systemS}{\textsl{BriskStream}\xspace}
\newcommand{\system}{\textsl{TStream}\xspace}
\newcommand{\lal}{\textsl{LOCK}\xspace}
\newcommand{\lwm}{\textsl{MVLK}\xspace}
\newcommand{\pat}{\textsl{PAT}\xspace}

\newcommand{\api}[1]{{{\fontfamily{cmtt}\selectfont{#1}}}}

\newcommand{\compact}{\vspace{-0pt}}
\newcommand{\subcompact}{\vspace{-0pt}}

\usepackage{pifont}

\usepackage{enumitem}   
\usepackage{tikz}

\begin{document}

\title{Towards Concurrent Stateful Stream Processing on Multicore Processors (Technical Report)}

\author{
\IEEEauthorblockN{Shuhao Zhang$^{1}$, Yingjun Wu$^2$, Feng Zhang$^3$, Bingsheng He$^1$}
\IEEEauthorblockA{$^{1}$National University of Singapore, $^{2}$Amazon Web Services, $^3$Renmin University of China}
} 


\maketitle

\input{abstract}
\compact
\input{introduction}
\compact
\input{background}
\compact
\input{motivation}
\compact
\input{system}
\compact
\input{detail}
\compact
\input{implementation}
\input{evaluation}

\compact
\input{related}

\compact
\input{conclusion}
\compact
\bibliographystyle{IEEEtran}
{
\bibliography{mybib}
}


\balance
\end{document}

%% file: abstract.tex
\begin{abstract}
Recent data stream processing systems (DSPSs) can achieve excellent performance when processing large volumes of data under tight latency constraints. 
However, they sacrifice support for concurrent state access that eases the burden of developing stateful stream applications. 
Recently, some have proposed managing concurrent state access during stream processing by modeling state accesses as transactions.
However, these are realized with locks involving serious contention overhead. 
Their coarse-grained processing paradigm further magnifies contention issues and tends to poorly utilize modern multicore architectures.
This paper introduces \system, a novel DSPS supporting efficient concurrent state access on multicore processors.
Transactional semantics is employed like previous work, but scalability is greatly improved due to two novel designs: 
1) dual-mode scheduling, which exposes more parallelism opportunities, 
2) dynamic restructuring execution, which aggressively exploits the parallelism opportunities from dual-mode scheduling without centralized lock contentions.
To validate our proposal, we evaluate \system with a benchmark of four applications on a modern multicore machine. 
The experimental results show that 
1) \system achieves up to 4.8 times higher throughput with similar processing latency compared to the state-of-the-art and
2) unlike prior solutions, \system is highly tolerant of varying application workloads such as key skewness and multi-partition state accesses.

\end{abstract}

%% file: introduction.tex
\vspace{+5pt}
\section{Introduction}
\label{sec:introduction}

The recent advances in data stream processing systems (DSPSs)~\cite{flink,Storm,heron,DBLP:conf/icde/WuT15} in terms of performance, elasticity, and scalability have accelerated their adoption in many emerging use cases. 
Modern stateful DSPSs such as Storm~\cite{Storm}, Heron~\cite{heron}, and Flink~\cite{flink} achieve high performance via disjoint partitioning of application states~\cite{Carbone:2017:SMA:3137765.3137777} -- often through key-based partitioning~\cite{partition_cost} so that each execution thread (i.e., executor) maintains a disjoint subset of states and thereby bypass the issue of \emph{concurrent state access}.
This type of design can lead to tedious implementation and ineffective performance in many cases (see later in section~\ref{subsec:example}). 

Several recent works propose to support \emph{concurrent state access} in stream processing, where large mutable application states may be concurrently accessed by multiple executors~\cite{acep,Affetti:2017:FIS:3093742.3093929}. 
State consistency is maintained by the system by adopting transactional semantics~\cite{acep,botan2012transactional}.
Specifically, the set of state accesses triggered by the processing of one input event at one operator is defined as a \emph{state transaction}. Multiple state transactions are concurrently executed using various \emph{concurrency control} mechanisms~\cite{acep,S-Store}.

Unfortunately, prior implementations are not free of bottlenecks when scaled up due to two reasons:
First, 
they are mostly built on centralized locking schemes, 
where every transaction has to access a set of monotonically increasing counters to decide if it is allowed to acquire locks of its targeting states. 
Despite its simplicity, 
it has serious contention issues and does not properly exploit the underlying multicore nature of modern CPU architectures.
Second, 
they commonly follow a coarse-grained processing paradigm, where an executor must finish all operations of processing one event before the processing of the next event can begin. 
This paradigm minimizes context switching overhead but overlooks opportunities for parallel processing.
In particular, the processing of one event may involve multiple conflict-free operations (e.g., stateless computation and multiple accesses to different states). Blocking one state access often unnecessarily blocks all operations of an event in this paradigm, further intensifying contention.

\begin{figure}
	\centering	
    \includegraphics[width=0.4\textwidth,bb=0 0 480 330]{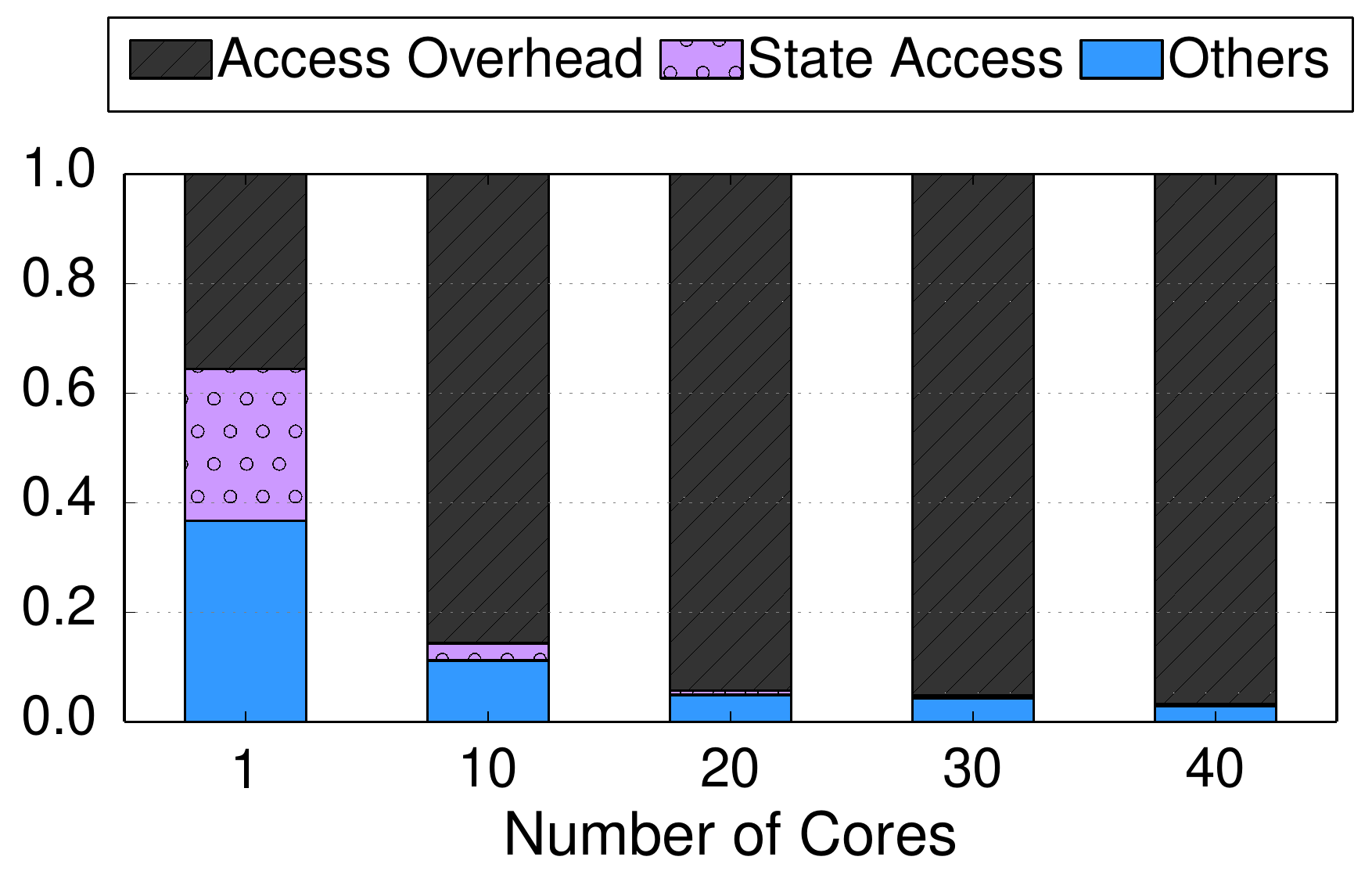}   	
	\caption{Severe lock contentions of the \pat scheme~\cite{S-Store}.}
	\label{fig:motivate} 
\end{figure}

Figure~\ref{fig:motivate} shows the evaluation results of the \pat scheme~\cite{S-Store}, the current state-of-the-art, on the Toll Processing~\cite{Linear} application.
We measure the average amount of time spent on 
(i) \emph{state access}, i.e., time spent accessing states, 
(ii) \emph{access overhead}, comprising of lock acquisition and blocking due to access contention, 
and (iii) \emph{others}, including all other operations (excluding state access) and overheads (e.g., context switching). 
As the number of cores used increases, the overhead of accessing states quickly dominates other operations due to serious contention. 
Therefore, we need a new solution for scaling concurrent state access in the DSPSs. 

This paper presents \system, a novel DSPS supporting efficient concurrent state access in the context of main memory multicore architectures.  
\system follows previous work~\cite{acep,botan2012transactional} of employing transactional semantics to handle concurrent state access but with much better scalability. 
The design of \system is inspired by our careful analysis of existing applications.
Stream processing usually consists of a set of operations that are repeated for every input event, 
and concurrent state access (if applied) often turns out to be a performance bottleneck.
This pattern guides us to abstract the processing as a \emph{three-step procedure}: preprocess, state access, and postprocess. 
While this formulation may appear to limit the flexibility of programming, it unlocks the potential for
simple and effective optimization opportunities, which we exploit to improve scalability.

\emph{First}, 
based on the three-step procedure,
we propose an execution strategy called \emph{Dual-Mode Scheduling}, which exposes more parallelism opportunities.
By carefully decoupling the second step (i.e., state access) from the processing logic, 
\system allows an executor to \emph{postpone} state access and instantly work on other input events without being blocked. 
Delaying state transactions to the last minute allows them to be processed in batches, enabling further optimizations when accessing state.

\emph{Second},
we propose a novel state transaction processing mechanism called \emph{Dynamic Restructuring Execution}. 
Specifically, \system restructures a batch of (postponed) transactions into a collection of sorted lists called \emph{operation chains}.
These can be evaluated in parallel without lock contention, significantly relieving contention overhead in concurrent state access.

\emph{In summary, we make the following contributions:}
First, we propose an efficient way of handling concurrent state access during stream processing with two novel designs.
Second, we implement the proposed designs as well as several state-of-the-art schemes in a fully functional DSPS~\cite{briskstream} optimized for multicore architectures.
We then compare them both theoretically and experimentally, revealing the scalability issues of prior solutions.
We open the full source code of the system and application benchmarks at \url{https://github.com/Xtra-Computing/briskstream/tree/TStream}.

%% file: background.tex
\section{PRELIMINARIES}
\label{sec:background}
\subcompact
\subsection{Data Stream processing} 
\label{subsec:example}

\begin{figure}[t]
    \captionof{table}{Summary of Terminologies}  
    \includegraphics[width=0.4\textwidth,bb=0 0 400 280]{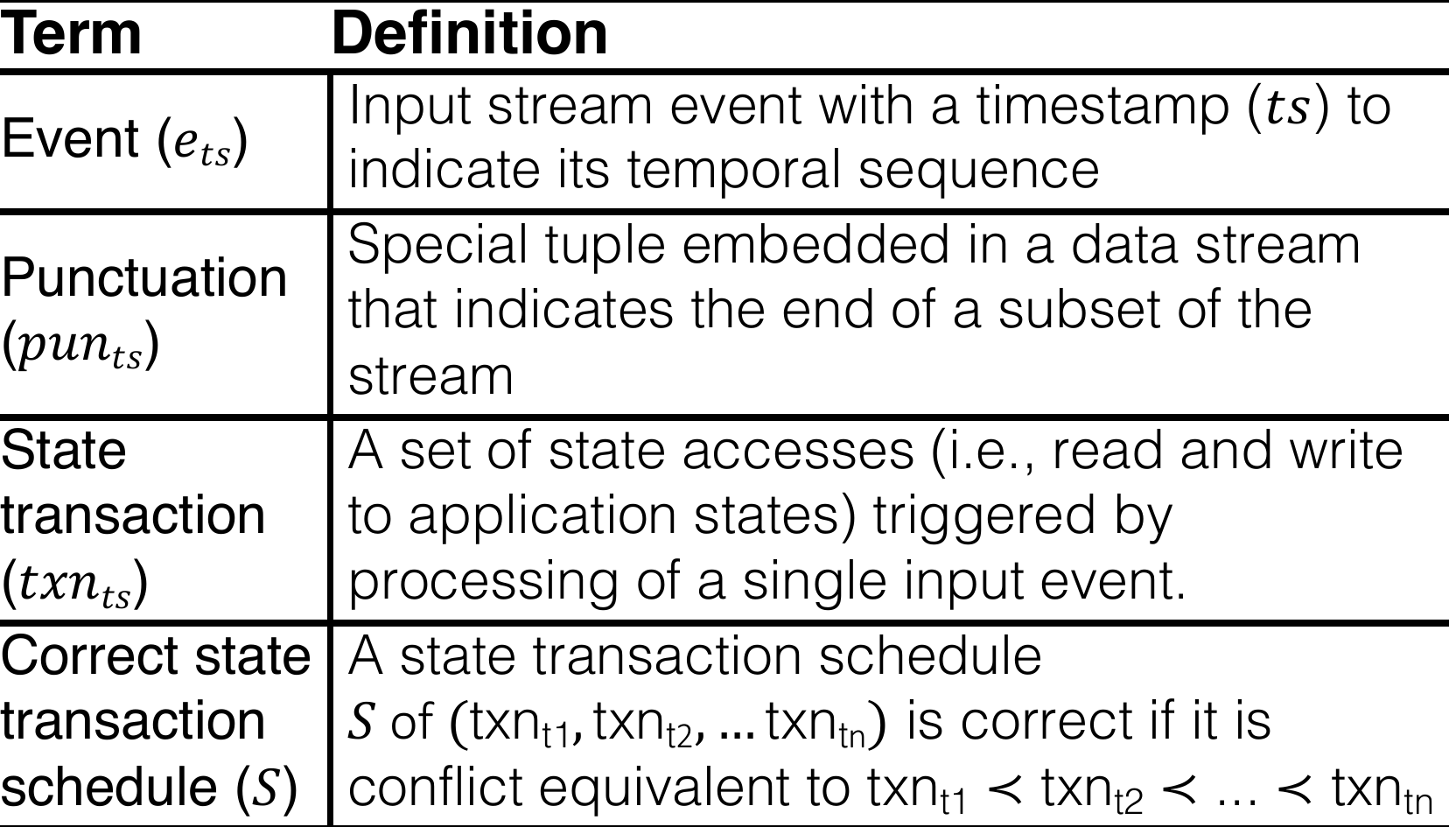}   
    \label{tbl:notations}  
\end{figure}

In this paper, we generally follow the definitions of data stream processing presented in~\cite{Aurora}, and we briefly recall them for completeness. 
We summarize the terminology used in this work in Table~\ref{tbl:notations}. 
Stream processing continuously processes one or more streams of \emph{events}.
Each event ($e_{ts}$) has a timestamp ($ts$) that indicates its temporal sequence. 
A streaming application contains a sequence of \emph{operators} that continuously process streaming events~\cite{profile}. 
To sustain a high input stream ingress rate, each operator may be spread across multiple \emph{executors} (e.g., Java threads), which handle multiple input events concurrently through stream partitioning~\cite{profile}. 
Operators often need to maintain states during processing for future reference~\cite{Carbone:2017:SMA:3137765.3137777}.  
\tony{To avoid state access conflict}, the common wisdom is adopting key-based stream partitioning~\cite{partition_cost} so that each executor maintains a \emph{disjoint subset} of states. 
Similarly, operators are required to maintain their states exclusively. 
To illustrate this, we use a simplified toll processing query (\textsl{TP}) from the Linear Road Benchmark~\cite{Linear} as an example.

\begin{figure*}[t]
\centering
\begin{minipage}{0.95 \textwidth}  
    \includegraphics[width=\textwidth,bb=0 0 650 150]{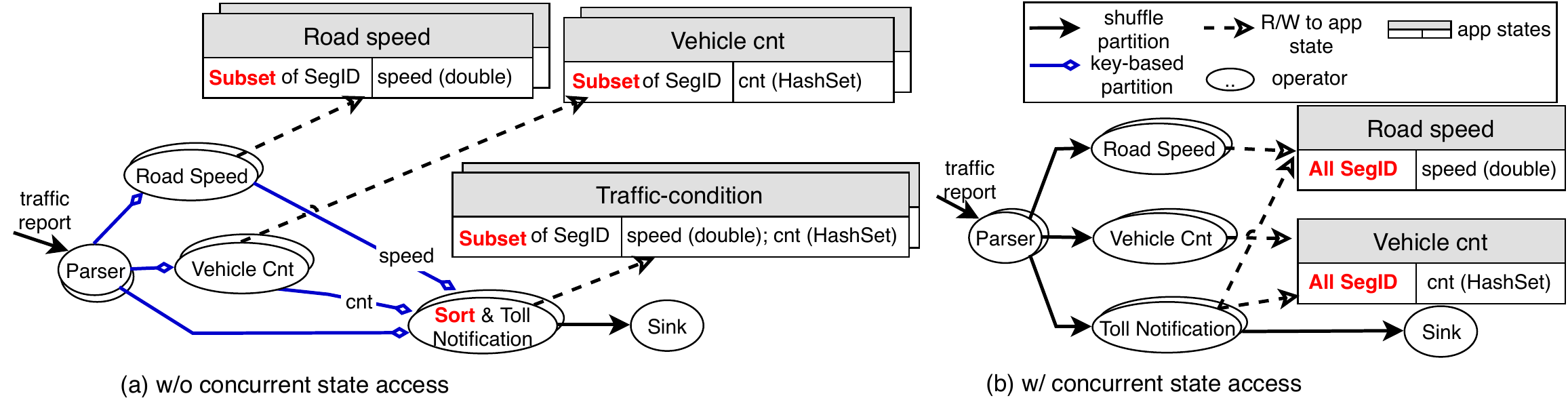}   
    \caption{Implementation of Toll Processing (\textsl{TP}).}   
    \label{fig:app}      
\end{minipage}    
\end{figure*}

\textbf{Motivating Example.}
\textsl{TP} calculates the toll every time a vehicle reports its position in a new road segment, in which tolls depend on the level of road congestion. 
It contains three key operators:
1) \texttt{Road Speed (RS)} computes average traffic speed of a road segment; 
2) \texttt{Vehicle Cnt (VC)} computes the average number of unique vehicles of a road segment;
3) \texttt{Toll Notification (TN)} computes the toll of a vehicle based on the traffic speed of and number of unique vehicles on the road segment where the vehicle is.

One common way~\cite{Aeolus} to implement \textsl{TP} is shown in Figure~\ref{fig:app} (a), where ovals denote operators and arrows denote data flow between operators.
\texttt{Parser} parses input events into traffic reports containing $<$timestamp, vehicle\_id, geo\_position, speed$>$, and the computed toll is continuously sent to \texttt{Sink} for output. 
Road congestion status (i.e., speed and count) are \emph{application states}~\cite{Carbone:2017:SMA:3137765.3137777}, which are maintained for future reference (the dashed arrows in the figure). 
To avoid state access conflict, a key-based partition scheme is adopted to split the input stream (the blue diamond arrows in the figure). 
It also prevents operators from concurrently accessing the same state by keeping their states exclusive. 

However, such an implementation can be tedious and ineffective.
\emph{First}, it requires users to carefully partition and sort the input stream by selecting appropriate keys. 
In this example, 
the application needs to ensure that any traffic report is processed only when \texttt{TN} receives the \emph{updated} road congestion status from \texttt{RS} and \texttt{VC}. 
Prior work~\cite{Aeolus} embeds tuple buffering and sorting operations (i.e., sort by vehicle id, geo-position and timestamp) inside the \texttt{TN} as highlighted in Figure~\ref{fig:app} (a). 
This manual approach is cumbersome and can lead to errors if tuples arrive too late (out of buffering limits).
\emph{Second}, 
the ineffectiveness stems from the duplication of large application states among operators. 
In this example, states maintained by \texttt{RS} and \texttt{VC} have to be repeatedly forwarded to \texttt{TN}.


\subcompact
\subsection{Concurrent Stateful Stream Processing}
\label{subsec:applications}
Many applications utilizing concurrent state access have been proposed covering various domains (e.g., Health-care~\cite{acep}, IoT~\cite{botan2012transactional}, and E-commerce~\cite{S-Store}). 
Despite the implementation differences, we identified three common application features.

\textbf{F1: Three-step procedure.} 
Each operator can be abstracted as a \emph{three-step procedure}: 
(1) preprocess input event (e.g. filter invalid input);
(2) accesses (shared) application states (e.g., read the road congestion status); 
finally, (3) perform further processing based on access results (e.g., compute toll based on road congestion status).

\textbf{F2: Determined read/write sets.}
Read/write sets of each state access is provided as arguments, which are inferred from the input event.
For example, which road segment to access is determined by the corresponding traffic report (i.e., key of state to access is tuple.geo\_position). 



\textbf{F3: Deterministic state access sequence.}        
State accesses must strictly follow their triggering event's timestamp sequence.
For example, computation of a toll must be processed according to the exact ``current" road congestion status to progress correctly, i.e., the toll should depend on neither \emph{stale} nor \emph{future} road congestion status. 

An implementation of \textsl{TP} utilizing concurrent state access is illustrated in Figure~\ref{fig:app} (b).
It contains the same operators, but road congestion information is shared among all operators and their executors. 
Particularly, congestion status for all road segments are maintained in two tables. 
One for the average road speed and the other for the count of unique vehicles. 
Input events are round-robin shuffled (the solid arrows in the figure) among all executors which concurrently process input events and access to the shared two tables.
\tony{Such an implementation significantly eased the burden of developing stateful stream application as developers do not need to manually split application states to ensure an exclusive and correctly ordered access among different threads.}

%

Unfortunately, concurrent state access introduces a new challenge of preserving consistency to DSPSs as multiple threads may concurrently access to the same application state with arbitrary sequence. 
Previous studies~\cite{acep,Affetti:2017:FIS:3093742.3093929,botan2012transactional,S-Store} advocate that concurrent state access can be efficiently managed with transactional semantics. 
We follow prior work~\cite{acep,botan2012transactional} and specifically adopt two key definitions.


\begin{definition}[State Transaction]
The set of state accesses triggered by processing of a single input event $e_{ts}$ at an operator is defined as one \emph{state transaction}, denoted as $txn_{ts}$. 
Timestamp ${ts}$ of a state transaction is defined as that of its triggering event. 
\end{definition}


\begin{definition}[Correct State Transaction Schedule]
\label{def:schedule}
A schedule of transactions $txn_{t1}$, $txn_{t2}$, ..., $txn_{tn}$ is correct if it is conflict equivalent to $txn_{t1}$ $\prec$ $txn_{t2}$ $\prec$ $...$ $\prec$ $txn_{tn}$. 
A DSPS ensuring a correct state transaction schedule always guarantees deterministic state access sequence (\textbf{F3}).
\end{definition} 

%% file: motivation.tex
\subsection{Existing Solutions Revisited}
\label{sec:motivation}


To ensure a correct schedule of concurrent state transactions,
various concurrency control mechanisms have been proposed. 
In the following, we revisit the representative ones. 

\subsubsection{\textbf{Lock-based approach (\lal)}}
An earlier study by Wang et al.~\cite{acep} described a strict two-phase locking (S2PL) -based algorithm that allows multiple state transactions to run concurrently.
To maintain a correct schedule (\textbf{F3}), 
it employs a \emph{lockAhead} process that compares each transaction's timestamp against a monotonically increasing counter to ensure that transaction with the smallest timestamp always obtains locks first, and hence guarantees proper state access sequence. 
By utilizing determined read/write sets (\textbf{F2}), once a transaction finished inserting its locks, the system can immediately increase the counter to allow next transaction to proceed without waiting for the transaction to finish processing.

\subsubsection{\textbf{Multiversion-Lock-based approach (\lwm)}}
To relax the rigorous lock incompatibility of the \lal scheme, Wang et al.~\cite{acep} propose to adopt multiversion concurrency control, where multiple copies of the same application state modified at different timestamps are kept by the system.
It further maintains a counter (called \texttt{lwm}) of each state to guard the state access order (\textbf{F3}).
Specifically, transactions need to compare their timestamp with the corresponding \texttt{lwm} counters before proceed (\textbf{F2}). 
A write is permitted only if the transaction's timestamp is equal to \texttt{lwm}; while a read is permitted as long as the transaction's timestamp is larger than \texttt{lwm} so that it can read a correct version of the state. 
During commits, a transaction needs to increase \texttt{lwm} of all its modified states.

\subsubsection{\textbf{Partition-based approach (\pat)}}
S-Store~\cite{S-Store} splits application states into multiple disjoint \emph{partitions}, and hence only needs to guard accessing order for transactions targeting the same partition by utilizing determined read/write sets (\textbf{F2}). 
During execution, each transaction needs to first compare its timestamp with monotonically increasing counters of its targeted partitions (maybe more than one) to ensure that it can proceed to insert locks (\textbf{F3}).
It is noteworthy that, despite being partitioned, two transactions can still conflict if their targeted partitions are overlapping. 
This is fundamentally different from key-based stream partitioning~\cite{partition_cost}.

\textbf{In Summary:}
There are two common scalability limitations in prior solutions. 
First, 
to ensure schedule correctness (\textbf{F3}), 
prior approaches compare the timestamp of every state transaction with a set of monotonically increasing counters to ensure that locks are granted in the desired order.
Despite its simplicity, such centralized locking schemes can have serious contentions, which would severely degrade  system performance.
Although \pat (i.e., S-Store) reduces such overhead when transactions access disjoint state partitions, it quickly devolves to \lal with more multi-partition transactions -- a common problem for partition-based approaches~\cite{Pavlo2012}.
Second, 
they all adopt a coarse-grained processing paradigm that sequentially evaluates the three-step procedure (\textbf{F1}) for each event; an executor (i.e., thread) must complete all operations of one event before starting next.
This minimizes potential context switching overhead, but overlooks parallelism opportunities and further intensifying contention. 



There are also many other existing concurrency control (CC) schemes~\cite{bernstein2009principles} that have not been applied to the problem of concurrent state access in stream processing. 
For example, the timestamp-ordering based (T/O) approach~\cite{Yu:2016:TTT:2882903.2882935,Yu:2014:SAE:2735508.2735511} is a popular CC technique that does not rely on locks.
In a T/O algorithm, each transaction is assigned a unique and monotonically increasing timestamp as the serial order for resolving conflicts.
Only ``fresh" transactions are allowed to proceed. For example, a read would be aborted if the state has been modified by a transaction with a larger timestamp. 
It appears that adopting a T/O algorithm to support concurrent stateful stream processing is straightforward, as we may simply assign a unique timestamp to each state transaction according to their triggering event. 
However, this could easily end up with livelock in situations where there are many executors, and eventually bring down system concurrency. 
For example, consider the system handling $txn_{t1}\texttt{=}read(x)$ and $txn_{t2}\texttt{=}write(x)$ in parallel and it is required that $txn_{t1}$ $\prec$ $txn_{t2}$. 
Suppose that the processing of $txn_{t2}$ is slightly faster and successfully modifies the write timestamp of $x$ (i.e., $W\_TS(x)$) to 2. 
Subsequently, $txn_{t1}$ will  be aborted and   never commit as its timestamp is now less than $W\_TS(x)$ (i.e., it comes too late). 
On the other hand, suppose that the system follows the original definition of T/O and assigns a larger timestamp to $txn_{t1}$, it would be committed in this case but state access order is violated.
As a result, neither way is able to ensure the correctness of the state transaction schedule.
Recent proposals of T/O algorithms generally follow the original T/O scheme but propose different timestamp allocation mechanisms~\cite{Yu:2016:TTT:2882903.2882935}, and hence will not help to resolve the issue. 
Other existing CCs (e.g., OCC~\cite{Wu2016}) are similarly not designed with an awareness of state access order (\textbf{F3}). 

%% file: system.tex
\section{\system Overview}
\label{sec:overview}

In this work, we follow prior work~\cite{acep,botan2012transactional} for employing transactional semantics on managing concurrent state access but propose two novel designs to better utilize multicore processors.
Those designs are largely inspired by the set of common features we identified 
from our careful analysis of existing applications discussed in Section~\ref{subsec:applications}.

\textbf{D1: Dual-Mode Scheduling (Exposing Parallelism).}
We propose an execution strategy called \emph{Dual-Mode Scheduling}, which exposes more parallelism opportunities. 
Instead of evaluating three steps (\textbf{F1}) sequentially for each input as done in the literature,
\system decouples the second step and postpones it to be evaluated later. 
As a result, \system has two modes: 1) the \texttt{compute} mode where executors continuously process more input events without being blocked due to state access; 2) the \texttt{state access} mode where executors collaboratively process a batch of postponed transactions with abundant parallelism opportunities. 


\textbf{D2: Dynamic Restructuring Execution (Exploiting Parallelism).}
We propose a novel \emph{Dynamic Restructuring Execution} strategy to efficiently evaluate a batch of transactions in the \texttt{state access} mode.
Leveraging {determined read/write sets} (\textbf{F2}), 
\system conceptually decomposes each state transaction into multiple operations, each targeting one state.
On top of that, with a {determined state access sequence} (\textbf{F3}), \system restructures those operations into timestamp-ordered lists (called operation chains), where one list is tied to one state and evaluated by one thread.
With this restructuring, operation chains can be evaluated in parallel, and 
state access conflict is avoided within the operation chains.

%% file: detail.tex
\section{Design Details}
\label{sec:design}

In this section, we discuss our designs in detail. 
We first describe \system's APIs for users to implement concurrent stateful stream processing applications. 
Then we discuss the implementation of dual-mode scheduling and dynamic restructuring execution.

\begin{algorithm}[t]
\footnotesize
  boolean $dualmode$;\tcp{\footnotesize{\color{Gray}flag of dual-mode scheduling}}
  Map $cache$;\tcp{\footnotesize{\color{Gray}thread-local storage}}
  \ForEach{\text{event $e$ in input stream}}
  {
    \uIf{$e$ is not punctuation\tcp{\footnotesize{\color{Gray}always true under prior schemes}}}{
        EventBlotter $eb$ $\gets$ \textbf{PRE\_PROCESS} ($e$);\tcp{\footnotesize{\color{Gray} e.g., filter events}}
        \textbf{STATE\_ACCESS} ($eb$);\tcp{\footnotesize{\color{Gray}issue one state transaction}}
        \uIf{dualmode}{
            \tcc{\footnotesize{\color{Gray}stores events whose state access is postponed under \system scheme.}}
      	    $cache$.add($<e$, $eb>$)\;
            
      	}\Else{
      	    \tcc{\footnotesize{\color{Gray}evaluates three steps contiguously under prior schemes.}}
      	    \textbf{POST\_PROCESS} ($<e$, $eb>$);\tcp{\footnotesize{\color{Gray}e.g., computes toll based on obtained road statistics}}
      	}
    }
    \Else{
        \tcc{\color{Gray}if the event is a punctuation, transaction processing can start.}
        \textbf{TXN\_START}()\tcp{\color{Gray}Triggers mode switching.}
        \ForEach{$<e$, $eb>$ $\in$ $cache$}{
            \textbf{POST\_PROCESS} ($<e$, $eb>$);
        }
    }
  }
  \caption{Code template of an operator}
  \label{alg:algo}
\end{algorithm}

\begin{algorithm}[t]
\footnotesize
 \KwIn{EventBlotter $eb$}
 \KwResult{$avg$}
 \Fn(\tcc*[h]{function as parameter}){\FRecurs{$ts$, key, speed}}{
    \textbf{READ}($SpeedTable$, $ts$, $key$, $eb$);\tcp{\footnotesize{\color{Gray} obtain average speed of a road segment, result is stored in $eb$.}}
    $avg$ $\leftarrow$ $eb.result$ \texttt{+} speed;\tcp{\footnotesize{\color{Gray} compute new average speed.}}    
    \Return $avg$\;      
  }

 \Begin{
	 \textbf{READ\_MODIFY}($SpeedTable$, $eb.ts$, $eb.key$, $eb.key$, $Fun$($ts$, $eb.key$, $eb.value$), $eb$);\tcp{\footnotesize{\color{Gray}update average speed of a road segment using return value of $Fun$.}}	
 }
   
 \caption{STATE\_ACCESS of \texttt{Road Speed}}    
  \label{alg:algo1}
\end{algorithm}

\begin{algorithm}[t]
\footnotesize
 \KwIn{EventBlotter $eb$}
 \KwResult{$cnt$}
 \Fn(\tcc*[h]{function as parameter}){\FRecurs{$ts$, key, vid}}{
    \textbf{READ}($CountTable$, $ts$, $key$, $eb$);\tcp{\footnotesize{\color{Gray}obtain hashset of CountTable.}}
	$eb.result$.insert(vid);\tcp{\footnotesize{\color{Gray}update the hashset.}}    
	$cnt$ $\leftarrow$ $eb.result$.size();\tcp{\footnotesize{\color{Gray}only unique vehicle counts.}}    
    \Return $cnt$\;      
  }

 \Begin{
	 \textbf{READ\_MODIFY}($SpeedTable$, $eb.ts$, $eb.key$, $eb.key$, $Fun$($ts$, $eb.key$, $eb.value$), $eb$);\tcp{\footnotesize{\color{Gray}update unique vehicle count of a road segment using return value of $Fun$.}}	
 }
   
 \caption{STATE\_ACCESS of \texttt{Vehicle Cnt}}    
  \label{alg:algo2}
\end{algorithm}
 
\begin{algorithm}[t]
\footnotesize
 \KwIn{EventBlotter $eb$}

 \Begin{
	 \textbf{READ}($SpeedTable$, $eb.ts$, $eb.key\_s$, $eb$);\tcp{\footnotesize{\color{Gray}obtain average speed of a road segment.}}
	 \textbf{READ}($CountTable$, $eb.ts$, $eb.key\_c$, $eb$);\tcp{\footnotesize{\color{Gray}obtain vehicle count of a road segment.}} 	   	
 }

 \caption{STATE\_ACCESS of \texttt{Toll Notification}}    
  \label{alg:algo3}
\end{algorithm}

\subcompact
\subsection{Programming APIs}
\label{subsec:query}
In line with many popular DSPSs, 
\system expresses an application as a DAG (Directed Acyclic Graph) with an API similar to that of Storm~\cite{Storm}. 
To support concurrent stateful stream processing, \system provides a list of user-implemented and system-provided APIs inside each operator. 
The former are user implemented based on their application requirements
and the latter function as library calls, 
similar to some existing frameworks~\cite{he2008mars}. 
Currently, all APIs are implemented in Java. 

\tony{User-implemented APIs are summarized in Table~\ref{tab:user_api}, which requires users to implement operations of an operator as a three-step procedure (\textbf{F1}). 
We leave full automation of this process for future work.}
A code template of an operator is shown in Algorithm~\ref{alg:algo}. 
\tony{State transaction is expressed through the \api{STATE\_ACCESS} API, which would be implemented by users using system-provided APIs.} 
Algorithm~\ref{alg:algo1}~\ref{alg:algo2}~\ref{alg:algo3} illustrate implementations of \api{STATE\_ACCESS} by using \texttt{Road Speed}, \texttt{Vehicle Cnt}, and \texttt{Toll Notification} as examples. 
All operations issued from one invocation of \api{STATE\_ACCESS} are treated as one state transaction. 

System-provided APIs are summarized in Table~\ref{tab:system_api}
\api{READ}, \api{WRITE}, and \api{READ\_MODIFY} stand for the atomic operation of a state transaction. 
For brevity, $table$, $timestamp$, and $EventBlotter$ arguments are omitted in Table~\ref{tab:system_api} and are shown in Algorithm~\ref{alg:algo2}. 
$Key$ and $Value$ stand for key and new value of the state to access, respectively. 
$opt$ means that the parameter is optional.
$Fun$ stands for a user-defined function such as increment by 1. 
$CFun$ stands for a user-defined function that determines whether the operation will be applied.  
Users can implement $Fun$ and $CFun$ by constructing system-provided APIs (e.g., a conditional update depends on a read operation), similar to the way of constructing \api{STATE\_ACCESS}. 
\api{TXN\_START} is used to indicate mode switching and is only used under \system's dual-mode scheduling scheme.

\begin{table}[]
\centering
\caption{User-implemented APIs}
\label{tab:user_api}
\resizebox{0.5\textwidth}{!}{%
\begin{tabular}{|p{3.2cm}|p{6cm}|}
\hline
APIs & Description \\ \hline
{\color{blue}EventBlotter} {\bf PRE\_PROCESS} ({\color{blue}Event}\ $e$)    
&  Implements the pre-process function (e.g., filter). It returns EventBlotter containing parameter values (e.g., read/write sets) extracted from $e$.           \\ \hline
{\color{blue}void} {\bf STATE\_ACCESS} ({\color{blue}EventBlotter}\ $eb$)    
& Implements the state transaction through constructing system-provided APIs such as $READ$, $WRITE$.        
\\ \hline
{\color{blue}void} {\bf POST\_PROCESS} ({\color{blue}Event}\ $e$, {\color{blue}EventBlotter}\ $eb$)    &  Implements the post-process function that is depended on results of state access (stored in EventBlotter).           \\ \hline
\end{tabular}%
}
\end{table}

\begin{table}[]
\centering
\caption{System-provided APIs}
\label{tab:system_api}
\resizebox{0.5\textwidth}{!}{%
\begin{tabular}{|p{4.2cm}|p{5cm}|}
\hline
APIs & Description \\ \hline
{\color{blue}void} \textbf{READ} ({\color{blue}Key} $d$, {\color{blue}EventBlotter}\ $eb$)  
&  issues a read request with key of $d$ and store results in $eb$ for further processing (i.e., post-process).          
\\ \hline
{\color{blue}void} \textbf{WRITE} ({\color{blue}Key} $d$, {\color{blue}Value} $v$, {\color{blue}opt} {\color{blue}CFun} $f^*$({\color{blue}Key} s))
& issues a modify request so that $state(d) \leftarrow v$ if $f^*(s)$ is true or $f^*(s)$ is null. If $d!\texttt{=}s$, this request involves data dependency.    
\\ \hline
{\color{blue}void} \textbf{READ\_MODIFY} ({\color{blue}Key} $d$, {\color{blue}Fun} $f$({\color{blue}Key} $t$), {\color{blue}opt} {\color{blue}CFun} $f^*$({\color{blue}Key} $s$))   
&  issues a read and modify request so that $state(d) \leftarrow f(t)$ if $f^*(s)$ is true or $f^*(s)$ is null.
\\ \hline
{\color{blue}void} {\bf TXN\_START} ()
&   triggers mode switching to process postponed transactions. 
\\ \hline
\end{tabular}%
}
\end{table}

\subcompact
\subsection{Dual-Mode Scheduling}
\label{subsec:SP}
As discussed in Section~\ref{sec:overview}, \system adopts a nonconventional processing strategy, where the state access step is postponed.
There are three key components to support such postponing efficiently and correctly: 
\emph{1) EventBlotter Maintenance} creates and initializes a thread-local auxiliary data structure called \emph{EventBlotter}, which acts as the data bridge linking the two processing modes; 
\emph{2) Processing Mode Switching} enables efficient and correct mode switching
in \system with punctuation technique; 
\emph{3) Progress Controller} generates punctuations and assigns timestamps to events and punctuations. \system requires punctuations to contain a monotonically increasing timestamp.


\subsubsection{EventBlotter Maintenance}
A key design decision in \system is to maintain a thread-local auxiliary data structure (implemented as a Java Class), called \emph{EventBlotter}, to track information (e.g., parameter values and processing results) of each postponed transaction. 
An EventBlotter is created by the system upon exiting \api{PRE\_PROCESS} (i.e., Line 5 of Algorithm~\ref{alg:algo})). 
Upon entering \api{STATE\_ACCESS} (i.e., Line 7 of Algorithm~\ref{alg:algo}), 
\system creates a state transaction with a list of \api{READ}, \api{WRITE}, \api{READ\_MODIFY} operations according to users' implementation. 
As mentioned before, state transaction is not instantly processed under \system's dual-model scheduling strategy. 
Instead, those operations are registered to \system to be evaluated later (Section~\ref{subsec:TP}). 
Their parameter values (e.g., read/write sets) are stored in the corresponding EventBlotter for future reference during transaction processing.
\api{POST\_PROCESS} might be required depending on the result of state accesses. 
To support this, we store input events and their corresponding EventBlotters in a thread-local map structure (i.e., Line 9 of Algorithm~\ref{alg:algo}), which will be processed after postponed transactions are processed.

\subsubsection{Processing Mode Switching}
\system relies on \emph{punctuation}~\cite{Tucker:2003:EPS:776752.776780} to periodically switch between two processing modes.
A {punctuation} is a special type of event that guarantees that no subsequent input event will have a smaller timestamp.
It is widely used in prior work for out-of-order stream processing~\cite{StreamBox,trill}.
Our usage of punctuation is different from the previous work~\cite{StreamBox}, as we target more fine-grained control at transaction processing rather than event processing.
Specifically, input events may be processed in arbitrary order in \system, but their issued transactions must be processed following a correct sequence (\textbf{F3}). 
\tony{A punctuation ensures that any state transaction issued before it should have a smaller timestamp than any ones issued after it.}
This sharply delineates the
timestamp boundary of a list of
transactions between any two consecutive punctuations
and 
gives \system hints  on how to effectively
process them.


To ensure the correctness of mode switching, 
\system artificially adds two barriers (via the \emph{Cyclicbarrier}~\cite{barrier}) to synchronize executors.
The \emph{first} barrier is added after the \api{TXN\_START} is called. 
This ensures EventBlotter maintenance for all events before the current punctuation is completed.
Only when all executors have switched to \texttt{state access} mode, can state access begin.
The \emph{second} barrier is added before the \api{TXN\_START} exits. 
This guarantees the correctness of the postprocessing step as executors do not resume to the \texttt{compute} mode until all postponed state accesses are fully processed (or aborted). 
\tony{
By processing transactions in batches, the overhead caused by these barriers will be amortized.
}

\begin{figure}[t]
\centering
     \includegraphics*[width=\textwidth,bb=0 0 850 150]{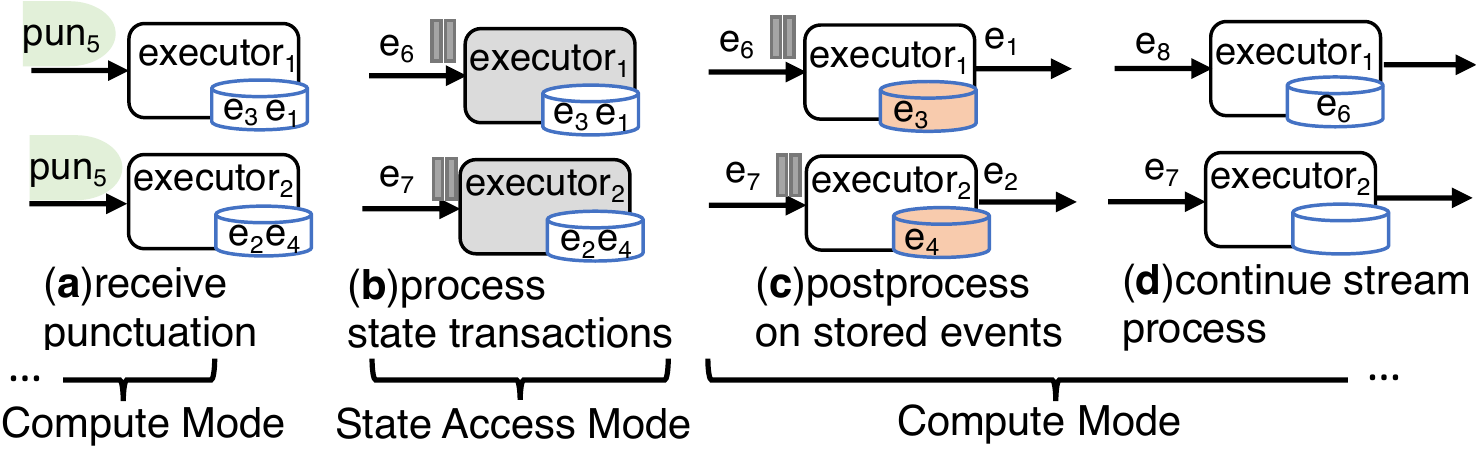}
     \caption{Example workflow of switching between modes.}                   
     \label{fig:punctuation}
\end{figure}

Figure~\ref{fig:punctuation} shows an example workflow of switching between modes:
(a) executors asynchronously switch to the \texttt{state access} mode when they receive punctuation with a timestamp of $5$ (i.e., Line 13 of Algorithm~\ref{alg:algo}) and no further input events are allowed to enter the system (e.g., $e_6$, $e_7$);
(b) subsequently, transaction processing is started (Section~\ref{subsec:TP}) once all executors are in the \texttt{state access} mode;
(c) when all postponed transactions are processed, executors are synchronously switched back to the \texttt{compute} mode to process (i.e., \api{POST\_PROCESS}) their stored \emph{unfinished} events, whose EventBlotter now contains the value of desired states (i.e., Line 14$\sim$15 of Algorithm~\ref{alg:algo});
finally, (d) executors are asynchronously resumed to process more input events.

%
 

\subsubsection{Progress Controller}
Punctuations are periodically broadcast to the input stream of each executor as done in~\cite{trill}. 
Punctuations' timestamp must monotonically increase to progress correctly, while events can have arbitrary timestamps as long as they are smaller than the next punctuation.
For simplicity, we assign both events and punctuation a monotonically increasing timestamp through the $fetch\&add$ instruction (via the \emph{AtomicInteger} in JDK8). 
This brings a minor impact on the overall performance as the system's bottleneck is on concurrent state access.

\subcompact
\subsection{Dynamic Restructuring Execution}
\label{subsec:TP}

The key problem in prior solutions (Section~\ref{sec:motivation}) is that all transactions are blocked while the one with the smallest timestamp is acquiring the locks it needs.
Such a coarse-grained scheme is simple to realize but introduces significant lock contention overhead.
We propose a fine-grained stream transaction execution mechanism, called \emph{dynamic restructuring execution}.
Specifically, \system restructures a batch of state transactions (obtained from dual-model scheduling) into a collection of operation chains that can be evaluated in parallel without any lock contentions.
It involves two key components: 
1) transaction decomposition, 
which breaks down each transaction into atomic operations, and inserting these into appropriate operation chains during the \texttt{compute} mode
and 2) transaction processing 
where the operation chains formed are evaluated in parallel
during the \texttt{state access} mode. 

\subsubsection{Dynamic Transaction Decomposition}
Once an event's EventBlotter is constructed and initialized, the executor is ready to postpone the issued transaction (i.e., Line 7 of Algorithm~\ref{alg:algo}). 
Conceptually, it decomposes each transaction into multiple state access operations, where each operation targets one application state. 
Then, it dynamically inserts decomposed operations into ordered lists (called \emph{operation chains}) with each list storing operations targeting one state (e.g., average road speed of one road segment). 
As the state transaction is expressed by constructing system-provided APIs, the decomposition is naturally achieved by treating one invocation of system-provided APIs (i.e., \api{READ}, \api{WRITE}, \api{READ\_MODIFY}) as an operation.
For example, two \api{READ} operations in Algorithm~\ref{alg:algo2} will be inserted into two operation chains as they target two different states from two tables.


Intuitively, any concurrent ordered data structure (e.g., self-balancing trees) can be used to implement the operation chain. 
However, inappropriate implementation can lead to large overhead in construction and processing.
We consider two properties of a suitable data structure.
\emph{First}, it must allow insertion from multiple threads simultaneously, while still guaranteeing the order of operations in the same chain.
\emph{Second}, it only requires sequential look-up rather than random access during processing.
Based on these considerations, we adopt the \emph{ConcurrentSkipList} due to its high insertion performance and small overhead compared to alternative designs, such as self-balancing trees, observed in prior work~\cite{pugh1989skip}.

\begin{figure}[t] 
\centering
      \includegraphics*[width=\textwidth,bb=0 0 650 100]{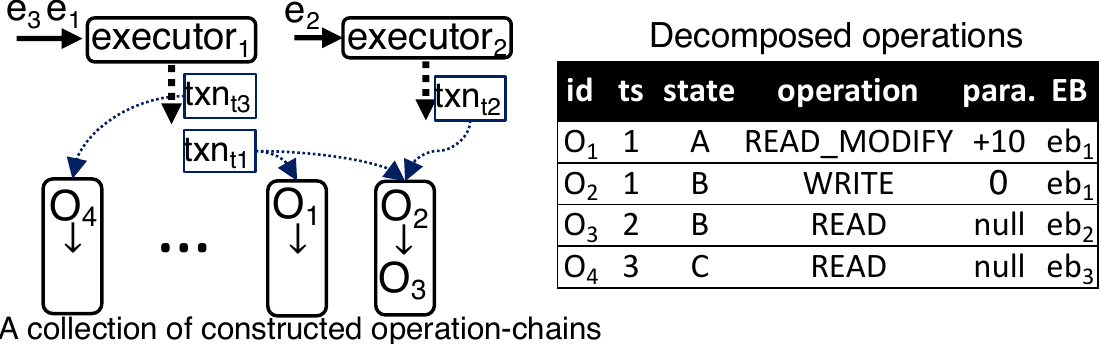}
     \caption{Transaction decomposition example.} 
     \label{fig:operationchains}
\end{figure}

Figure~\ref{fig:operationchains} illustrates the decomposition process for three transactions.
$txn_{t1}$ is decomposed into two operations, $O_1$ and $O_2$. 
Each operation is annotated with timestamp ($ts$) of its original transaction, targeted state ($state$), access operation ($operation$), and parameters ($para.$) including read/write sets and dependent functions. 
$O_2$ and $O_3$ are inserted into the same operation chain as they target the same state $B$. 
As $O_2$ has a smaller timestamp than $O_3$, the chain is sorted as $O_2\rightarrow O_3$.
$O_1$ and $O_4$ form another two chains as they target different states. 
Note that EventBlotters ($EBs$) are also embedded in the operation (i.e. the last column of the table in Figure~\ref{fig:operationchains}) so that they can be tracked during transaction processing for recording access results.

\subsubsection{Parallel Transaction Processing}
When executors are all switched to the \texttt{state access} mode, they proceed to collaboratively process the formed operation chains. 
There are two cases that we need to consider.
\begin{myitemize}
\item \emph{Case 1:} 
there are no data dependencies among different operation chains. 
Then, one executor simply sequentially walks (i.e., evaluate) through an operation chain from the top (i.e., operation with the smallest timestamp). 
All operation chains can be processed in parallel by multiple executors without any contentions. 
\item \emph{Case 2:} 
there are data dependencies among different operation chains. 
For example, a write operation of one state is dependent on a read operation of another state. 
\system handles data dependencies with a simple yet effective iterative process.
\end{myitemize}

\textbf{Handling Data Dependency.}
During transaction decomposition, \system records dependency information of operation chains, 
\tony{e.g., $chain_A$ depends on $chain_B$ if there is at least one operation targeting state $A$ and dependent on state $B$.} 
This dependency recording process is lightweight (i.e., simply marking $chain_A$ during operation insertion) without contentions.
During transaction processing, \system first parallelly process those operation chains with no data dependencies.
Then it parallelly processes the remaining ones which are dependent on those previously processed.
This iterative process continues until all operations are processed. 

This design has low overhead at tracking data dependencies (only at operation chain level) with high parallelism, but some operations may be processed out-of-order: 
operations with larger timestamp without dependencies on others may be processed earlier.
To handle this issue, \system maintains multiple versions (i.e., updated by operations with different timestamps) of a state during the processing if there are dependencies on it.
This ensures that subsequent reads will get the correct version (i.e., not necessarily the latest one) of the targeted states.
After the current batch of transactions is processed, all versions of a state except the latest are expired and can be safely garbage collected, restoring all states to having only a single version. 

The amount of memory used for keeping multiversions of states is related to the punctuation interval. 
Suppose there are $N$ transactions with a unique timestamp to handle between two subsequent punctuations and each transaction touches $m$ states with a size of $s$. 
Then, the amount of memory required to preserve different versions of the shared states is up to $N * m * s$. 
The upper limit is hit only if all transactions touch the same set of states (i.e., each state will be touched by $N$ times) and all operations depend on each other (i.e., every state needs to maintain multiple versions).
Taking \textsl{SL} as an example and suppose punctuation interval is 500, as each transaction touches up to four different states and each state has a size of 100 bytes (Section VI), the memory required is hence up to 500 * 4 * 100 = 200 Kbytes. 

This memory usage is stable during processing.  
As mentioned in Section IV-B, multiversions of a state except the newest version can be garbage collected immediately after \system switches back to \texttt{compute} mode. 
This is because the subsequent transactions will never need to access the previous versions of states except the latest ones as they are guaranteed to have larger timestamps. 

\textbf{Handling Transaction Abort.}
\tony{
If an update violates state consistency (e.g., road speed can not be negative), it has to be aborted, causing the corresponding transaction to be aborted. 
\system marks the corresponding input event as ``rejected'' to notify users via the output stream.
Due to the synchronization barriers, \system can abort transactions by simply removing them (i.e., skip the offending update operation during processing) from the batch of transactions to process before resuming the \texttt{compute} mode. 
Application semantics hence do not change under different schemes (i.e., \lal, \lwm, \pat and \system) and how state accesses are executed or aborted is transparent to users, who only know if the event is successfully executed or rejected. 
Note that the goal of \system is to support concurrent stateful stream processing rather than supporting arbitrary user-defined transaction aborts.
}

\subcompact
\subsection{Consistency Guarantee}
\label{subsec:consistency}
We now discuss how \system ensures that concurrent state transaction processing preserve the following properties~\cite{Affetti:2017:FIS:3093742.3093929} in order to provide state consistency guarantee:

\begin{itemize}
\item \textbf{Atomicity} requires all or none operations of a state transaction to be executed. Although transactions are restructured under TStream, all operations between two subsequent punctuations must be either executed or aborted before TStream can resume to continue processing new input events (i.e., new transactions). 
Hence, transactions between two subsequent punctuations will be executed (or aborted) and atomicity is always satisfied in TStream.
\item \textbf{Consistency} requires application state being valid (e.g., average road speed must be larger than 0) after all updates applied. TStream processes all operations of one operation chain (targeting at one state) in one thread sequentially. 
Once a write operation violates consistency, it will be aborted and subsequently abort the corresponding transaction. 
Hence, TStream can always satisfy consistency.
\item \textbf{Isolation} requires concurrent transactions are executed as if they are executed in some sequential order. This is naturally guaranteed as TStream always guarantee correct state transaction schedule (Definition 2): transactions are executed as if they are executed follows event sequence.
\item \textbf{Durability} requires modification to state are durable. 
TStream can replicate states stored in memory to disk before resuming to \texttt{compute} mode to satisfy durability. In this paper, we assume states are always kept in main memory.
\end{itemize}

\subcompact
\subsection{System Optimizations}
\label{subsec:opt}

\textbf{Transaction Batching.}
\system focuses on achieving a reasonable latency level with high throughput. 
Compared to the existing approaches, \system does not instantly process each issued state transaction but periodically processing batches of state transactions.
The interval size of two subsequent punctuations hence plays an important role in tuning system throughput and processing latency. 
If a large interval is configured, the system waits for a longer period before processing transactions, which increases worst-case processing latency. 
This is because some events are waiting (i.e., stored on its executor) for their issued transactions to be processed. 
Conversely, a small interval size might drop system throughput due to insufficient parallelism to amortize synchronization overhead.
We will evaluate the effect of the punctuation interval in our experiments.

\textbf{NUMA-Aware Processing.}
\label{subsec:NUMA_OPT}
Following previous work~\cite{Porobic:2012:OHI:2350229.2350260,6816692}, 
we consider three different design options for processing operation chains on multisocket multicore architectures.
\emph{1) Shared-nothing:}
we maintain a pool of operation chains per core. 
Essentially, decomposed operations are dynamically routed to predefined cores by hash partitioning.
One executor is responsible for processing all operation chains in one core.
This configuration minimizes cross-core/socket communication during execution but it may result in workload imbalance;
\emph{2) Shared-everything:} 
we maintain a centralized pool of operation chains, which is shared among all executors;
\emph{3) Shared-per-socket:}
we maintain a pool of operation chains per socket. 
Executors of the same socket can thus share their workloads, but not across different sockets. 

Workloads are shared among multiple executors under shared-everything and shared-per-socket configuration. 
Instead of statically assigning tasks to each executor, dynamic {work-stealing}~\cite{Blumofe:1999:SMC:324133.324234} can be applied to achieve better load balancing.
Specifically, multiple executors (in the same sharing group) continuously fetch and process an operation chain as a task from their shared task pool. 
Such a configuration achieves better workload balancing but pays more for cross-core (and cross-socket in the case of the shared-everything configuration) communication overhead compared to the shared-nothing configuration.
We will evaluate \system with varying NUMA-aware processing configurations in our experiments.


\subcompact
\subsection{System Limitations}
\label{subsec:limit}
\system relies on a \emph{mostly single-version} concurrency control (i.e., multiversion of a state is maintained if there are dependencies on it) without any centrally contented locks via two novel designs (\textbf{D1} and \textbf{D2}).
However, it has two main limitations.
First, 
\system performs the best when there are no data dependencies among operation chains in the workload (e.g., \textsl{TP}) as all operation chains can be processed in parallel. 
In our experiments, we show that \system can still perform better compared to previous solutions when the workload contains a lot of data dependencies (e.g., \textsl{SL}) owing to the unlocked parallelism opportunities.
Second, 
\tony{
\system pays high overhead when aborting multi-write transactions. 
This is because \system decomposes each transaction into the most fine-grained pieces (i.e., operations) and distributes them into multiple (could be many) operation chains in order to enlarge parallelism opportunities. 
Subsequently, the abortion of a  multi-write  transaction may roll back multiple operation chains. 
We plan to adopt an optimistic execution strategy~\cite{Wu2016} to further enhance our system in future work. 
}

%% file: implementation.tex
\section{Implementation Details}
\label{sec:impl}

\system adopts a modular design. It contains two modules:
1) The stream module is based on \systemS~\cite{briskstream}, a highly optimized general purpose DSPS with an architecture similar to Storm. 
\tony{We extend its original APIs as discussed in Section~\ref{subsec:query} to support concurrent statful stream processing;}
2) The state module is based on the Cavalia~\cite{Wu2016} database, which implements the system-provided APIs for managing state transaction execution.
Our proposed techniques can be generalized to other existing DSPSs, such as Storm and Flink, by integrating the state module into other DSPSs with minor efforts.
However, our solution is mainly designed for the shared-memory multicore environment.
It might require a system redesign to fully take advantage of the design and implementation of \system in a distributed environment such as Flink/Storm~\cite{profile}.

\system does not rely on key-based partitioning~\cite{partition_cost} as executors are allowed to access \emph{any} part of application states. 
This allows \system to fuse~\cite{Hirzel:2014:CSP:2597757.2528412} operators into a single joint operator to eliminate the impact of cross-operator communication, which is known to be a serious performance bottleneck of DSPSs~\cite{briskstream,profile,Zeuch:2019:AES:3303753.3316441}. 
For example, \texttt{Road Speed}, \texttt{Vehicle Cnt}, and \texttt{Toll Notification} operator are fused into one joint operator. A switch-case statement is used to invoke the corresponding operator logic for each input event.
Subsequently, \system allows this joint operator to be scaled to any number of executors without violating the consistency of state. 
Input events can be round-robin shuffled among all executors of the joint operator to ensure load-balancing.
This further simplifies application development and reduces the complexity of execution plan optimization~\cite{briskstream}.

%% file: evaluation.tex
\section{Evaluation}
\label{sec:eva}
In this section, we show that \system manages to better exploit hardware resources compared to the state-of-the-art by a detailed experimental evaluation.

\subcompact
\subsection{Benchmark Workloads}
A benchmark for transactional stream processing is still an open problem. 
Previous work~\cite{botan2012transactional, acep, S-Store} typically chooses a couple of applications in an ad hoc manner to evaluate their system's performance.
For our experiments, we follow the four criteria proposed by Jim Gray~\cite{Gray:1992:BHD:530588} and assemble four applications: \emph{Grep and Sum (\textsl{GS})}, \emph{Streaming Ledger (\textsl{SL})}, \emph{Online Bidding (\textsl{OB})}, and \emph{Toll Processing (\textsl{TP})}.

We briefly describe how our chosen applications achieve the four criteria: 
1) Relevance: 
    the applications cover diverse runtime characteristics and types of state access;
2) Portability: 
    we describe the high-level functionality of each application and note that these can be ported easily
    to other DSPSs supporting concurrent state access;
3) Scalability: 
    the applications chosen can be configured with different sizes;
4) Simplicity: 
    the applications are chosen with simplicity in mind so that the benchmark is understandable. 
 
Our benchmark covers different aspects of application features. 
\emph{First}, 
our applications cover varying runtime characteristics. 
Specifically, 
when a single core is used,
\textsl{TP} spends 39\% of the total time in \texttt{compute} mode, and this ratio is 29\% and 22\% for \textsl{SL} and \textsl{OB}, respectively. \textsl{GS} spends relatively less time in \texttt{compute} mode (13\%), and more time in \texttt{state access} mode.
\emph{Second}, 
they cover different types of state transactions. 
Specifically, 
\tony{different combinations of \api{READ}, \api{WRITE} and \api{READ\_MODIFY} operations are involved in the issued state transactions from different applications.}
Furthermore, \textsl{SL} has heavy data dependencies when handling transfer requests, i.e., updating one user account requires a read of another user account.

We have described \textsl{TP} earlier in Figure~\ref{fig:app} (b) in Section~\ref{sec:background}. 
Now, we describe the remaining applications, \textsl{GS}, \textsl{SL}, and \textsl{OB} including its application scenario, implementation details, and input setup. 
In all applications, we use a \texttt{Parser} operator to generate and parse input events and feed the remaining operators and a \texttt{Sink} operator to measure system performance. 
\tony{All applications need to maintain shared mutable states among operators, and concurrent state accesses (modelled as state transactions) shall follow a correct schedule (Def~\ref{def:schedule}, Section~\ref{sec:background}).}



\emph{\underline{Grep and Sum (GS)}:}
\textsl{GS} represents a synthetic scenario
where an application needs to read or 
update large shared mutable states 
and subsequently perform a computation based
 on    the obtained state values.
 \begin{wrapfigure}{r}{0.24\textwidth}
 \centering	
 	\includegraphics*[width=0.24\textwidth]{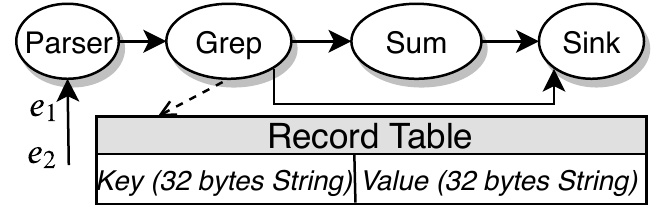}   
 	\caption{Grep and Sum (\textsl{GS}).}
 	\label{fig:gs} 
 \end{wrapfigure}
\texttt{Grep} issues a state transaction to access a list of records for each input event.
If an event triggers a state transaction with a list of \api{READ} operations, 
\texttt{Grep} forwards the input event with the returned state values to \texttt{Sum}; 
otherwise, it updates the state with a list of \api{WRITE} operations and forwards the input event to \texttt{Sink}.
\texttt{Sum} performs a summation of the returned state values from \texttt{Grep}. 
After \texttt{Sum} finishes its computation, it emits the result as one event to \texttt{Sink}. 
A table of 10k unique records is shared among all executors of \texttt{Grep}. 
Each record has a size of $\sim$128 bytes including JVM reference overhead, and 
each transaction length is 10 (i.e., ten accesses per transaction).
 
\emph{\underline{Streaming Ledger (SL)}:}
\textsl{SL} is suggested by a recent commercial DSPS, Streaming  Ledger~\cite{Transactions2018}.
\begin{wrapfigure}{r}{0.27\textwidth}
\centering	
	\includegraphics*[width=0.27\textwidth]{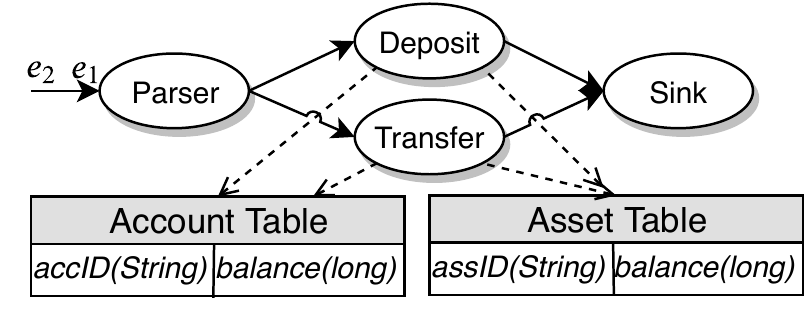}   
	\caption{Streaming Ledger (\textsl{SL}).}
	\label{fig:sl} 
\end{wrapfigure}   
It processes events that involve wiring money and asset between accounts.
The detailed descriptions are omitted here for brevity and can be found in the white paper~\cite{Transactions2018}.
\texttt{Deposit} processes requests that top-up user accounts or assets. 
\texttt{Transfer} processes requests that transfer balances between user accounts and assets. 
The updating results (success/fail) are passed to \texttt{Sink}.
The account and asset tables (each containing 10k unique records) are shared among all executors of \texttt{Deposit} and \texttt{Transfer}.
Each record has a size of $\sim$100 bytes including JVM reference overhead.
Transaction length is four for transfer request (i.e., transferring from a pair of account and asset to another pair) and is two for deposit request (i.e., update a pair of account and asset). 
We set a balanced ratio between transfer and deposit requests (i.e., 50\% each) in the input stream. 
 
\emph{\underline{Online Bidding (\textsl{OB})}:}
\textsl{OB} represents a simplified online   
\begin{wrapfigure}{r}{0.24\textwidth}
\centering	
	\includegraphics*[width=0.24\textwidth]{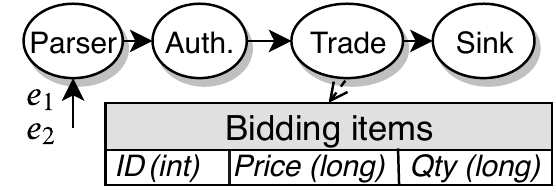}   
	\caption{Online Bidding (\textsl{OB}).}
	\label{fig:ob} 
\end{wrapfigure}
bidding system~\cite{bidding}. 
\texttt{Auth} 
authenticates trade requests and dispatches valid requests for further processing. 
\texttt{Trade} handles three types of requests including 
(1) \emph{bid request} reduces the quantity of its requested item if the bid price is larger or equal to the asking price and otherwise rejected. 
If the item has insufficient quantities, the bid request is also rejected. 
(2) \emph{alter request} modifies the prices of a list of requested items. 
(3) \emph{top request} increases the quantity of a list of items. 
The ratio of bid, alter, and top requests is configured as 6:1:1. 
A table of 10k unique bidding items are shared among all executors of \texttt{Trade}. 
Each record has a size of $\sim$50 bytes including JVM reference overhead.
The transaction length of both the alter and the top request is 20, and that of the bid request is one. 

\emph{\underline{Toll Processing (TP)}:}
In this work, we focus on evaluating mechanisms to support concurrent state access during stream processing.
We hence evaluate the implementation utilizing concurrent state access as illustrated previously in Figure~\ref{fig:app}(b) and omit the discussion of the conventional implementation. 
Each record in the road speed table has a size of $\sim$80 bytes, and record size in the vehicle count table varies depending on the number of items in the HashSet, i.e., $\sim$32$*$(2+$|items|$) bytes. 
State transactions from \texttt{Road Speed} and \texttt{Vehicle Cnt} has a length of one (i.e., update one record from one table) and those from \texttt{Toll Notification} has a size of two (i.e., read one record from the two tables).

%
\subcompact
\subsection{Experimental Setup}
We conduct all experiments on a 4-socket Intel Xeon E7-4820 server with 128 GB DRAM. The OS kernel is Linux 4.11.0-rc2.
Each socket contains ten 1.9GHz cores and 25MB of L3 cache and is connected to the other three sockets via Intel QPI.
NUMA characteristics, such as local and inter-socket idle latencies and peak memory bandwidths, are measured with Intel Memory Latency Checker~\cite{mlc}. 
Specifically, local memory latency (ns) is 142.6 and remote is 327.5, and local memory bandwidth (MB/sec) is 20564.8 and remote is 9944.
The number of cores assigned to the system, the size of the punctuation interval and NUMA-aware processing strategies are system parameters that can be varied by users. 
We vary both parameters in our experiments. 
We use a punctuation interval of 500 and shared-nothing processing as the default execution configuration.
We pin each executor on one core and assign 1 to 40 cores to evaluate the system scalability. 

Application states are randomly populated and evenly distributed to each executor before execution and are kept the same among different tests.
To present a more realistic scenario, we model the access distribution as Zipfian skew, where certain states are more likely to be accessed than others. 
For \textsl{GS}, \textsl{SL}, and \textsl{OB}, we set the skew factor to 0.6. 
For \textsl{TP}, we use the datasets from the previous work~\cite{profile}, which accesses 100 different road segments with a skew factor of 0.2. 
Application states may be partitioned beforehand and a multi-partition transaction will access multiple partitions.
Unless explicitly mentioned, we set the length and ratio of multi-partition transactions as 4 and 25\%, respectively. 
That is, each multi-partition transaction will access four different partitions, and 25\% of all transactions are multi-partition transactions. 
\texttt{Toll Notification} of \textsl{TP} accesses one record from two tables, and it hence always accesses two partitions.
We use a punctuation interval of 500 and shared-nothing as the default execution configuration.

We implement three competing schemes including the lock-based approach (\lal)~\cite{acep}, multiversion-lock-based approach (\lwm)~\cite{acep}, and partition-based approach (\pat)~\cite{S-Store} into \system. 
The original implementation of \lal and \lwm~\cite{acep} is not public available and the pseudocode given in the original paper is a single-threaded program. 
As also pointed out by the original paper~\cite{S-Store} Section 4, the implementation of S-Store uses a single-core for shared state accesses on one node. 
%
We hence adopt similar ideas from prior works and re-implement the corresponding multi-threaded version into \system to utilize multicore environment. 
We also examine the system performance when locks are completely removed from the \lal scheme, which is denoted by \textsl{No-Lock}, representing an upper bound on the system performance.

\textbf{Evaluation Overview.} 
We first show the overall performance comparison of different schemes on the benchmark suite (Section~\ref{subsec:usecases}).
Next, we provide transaction processing time breakdown for different schemes using \textsl{SL} as an example (Section~\ref{subsec:breakdown}). 
Then we evaluate \system under varying workload configurations (Section~\ref{subsec:single}). 
Finally, we perform a sensitivity study of \system in Section~\ref{subsec:factor} to validate the efficiency of our reimplmentation.

\subcompact
\subsection{Overall Performance Comparison}
\label{subsec:usecases}
\noindent
\framebox{
    \parbox{\dimexpr\linewidth-4\fboxsep-4\fboxrule}{
        \textbf{Finding (1):}        
        \system outperforms prior schemes by up to 4.8 times while ensuring a correct transaction schedule for all applications at large core counts.
    }
}

\begin{figure}[t]
    \centering
     \begin{center}
     \fbox{
     \includegraphics[width=0.8\columnwidth]
         {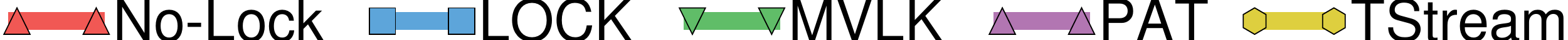}
     }
    \end{center}        
    \captionsetup[subfigure]{font=scriptsize,labelfont=scriptsize}
    \subfloat[\textsl{GS} (50\% read requests)]{
        \includegraphics[width=0.38\textwidth]
            {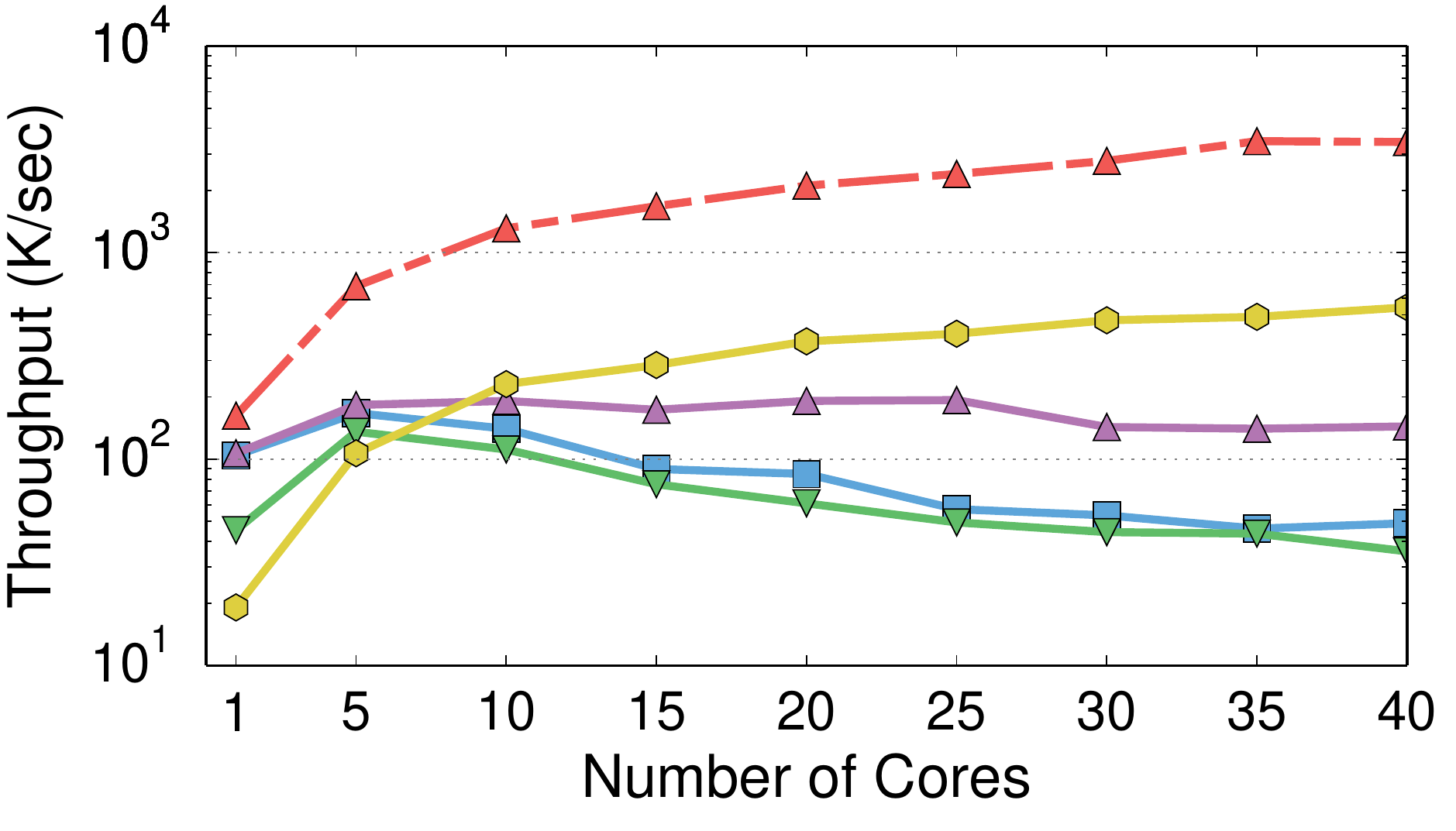}
    } 
       
    \subfloat[\textsl{SL}]{
        \includegraphics[width=0.38\textwidth]
            {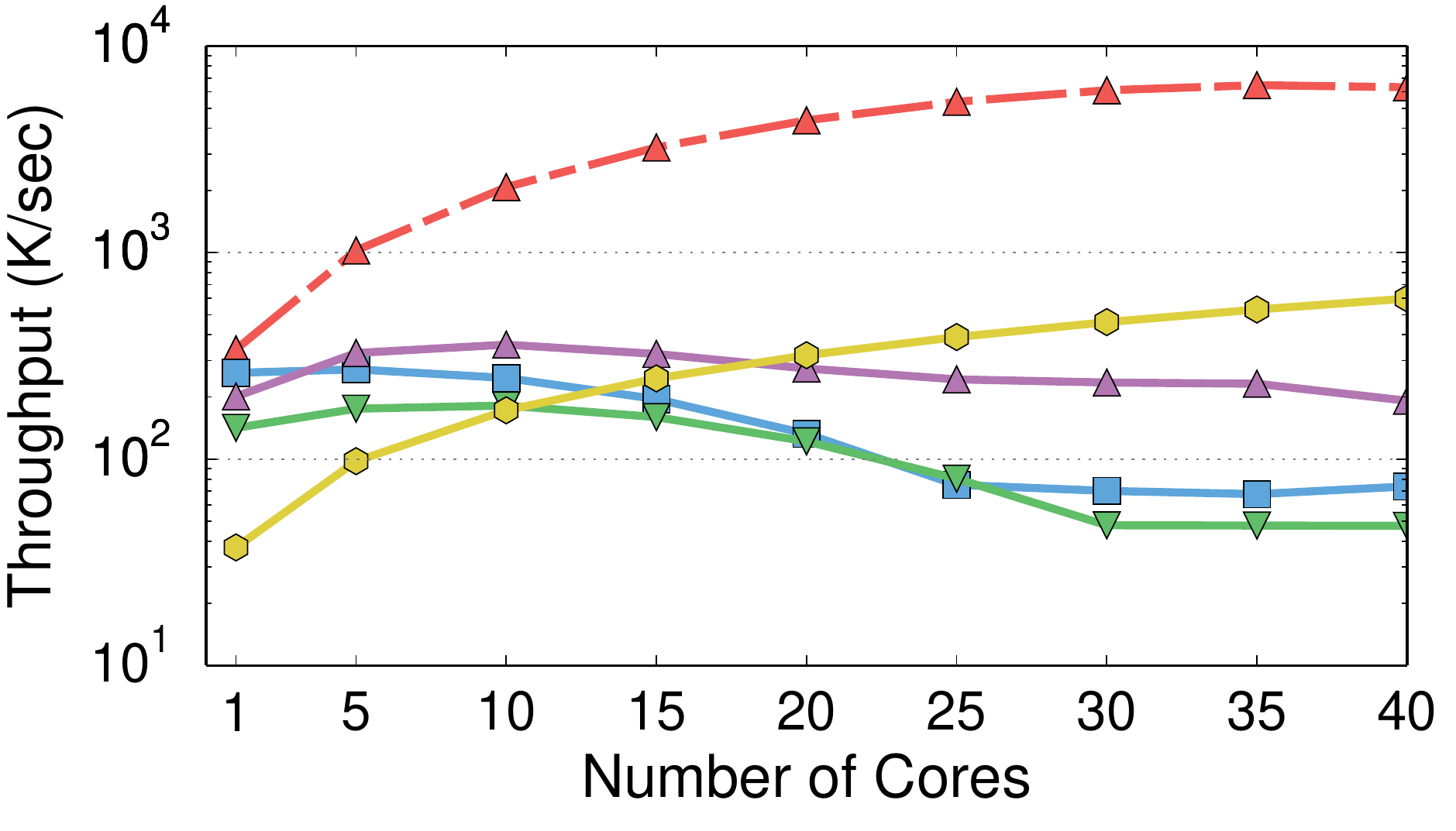}
    }       
    
    \subfloat[\textsl{OB}]{
        \includegraphics[width=0.38\textwidth]
            {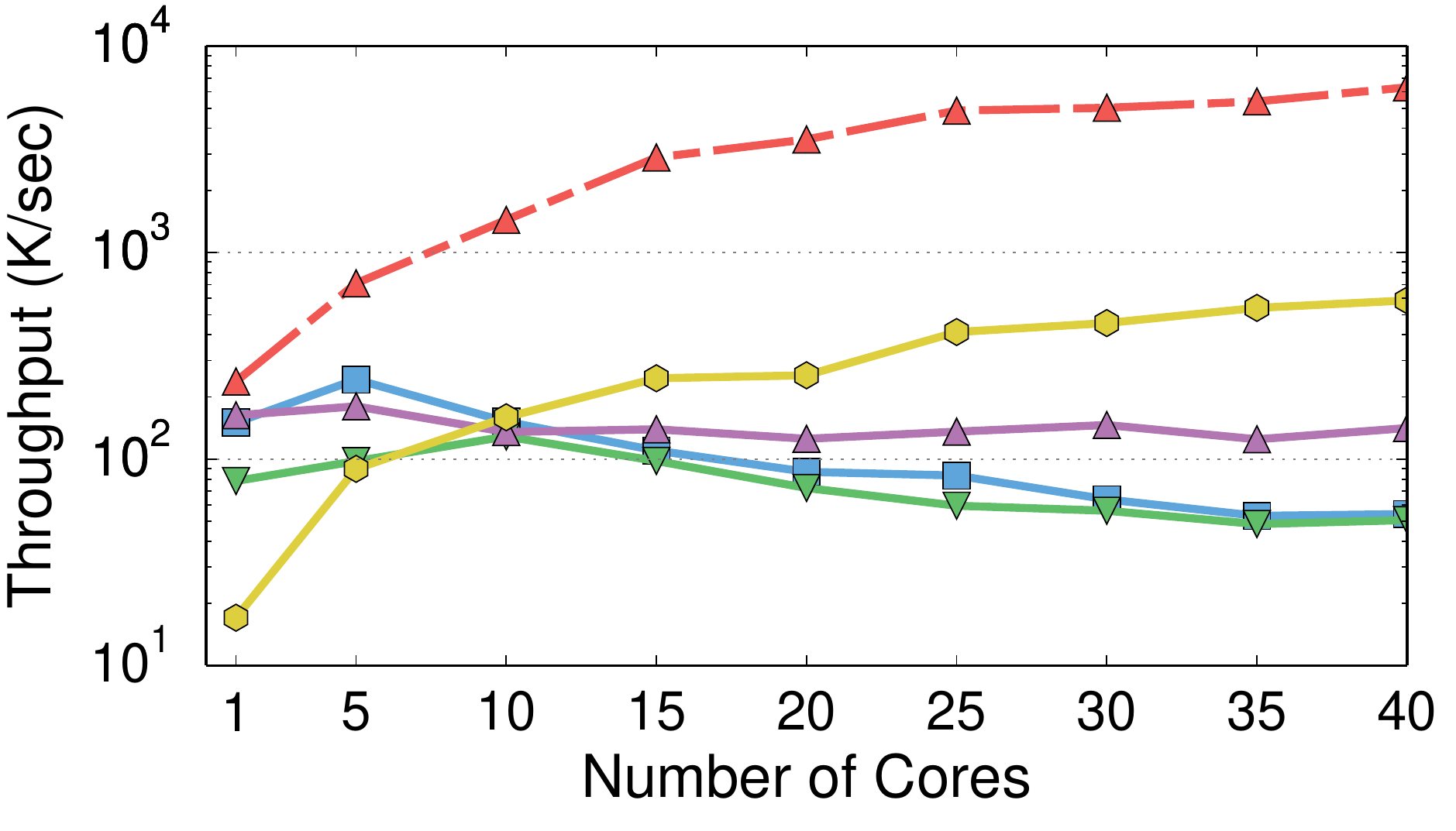}
    } 
    
    \subfloat[\textsl{TP}]{
        \includegraphics[width=0.38\textwidth]
            {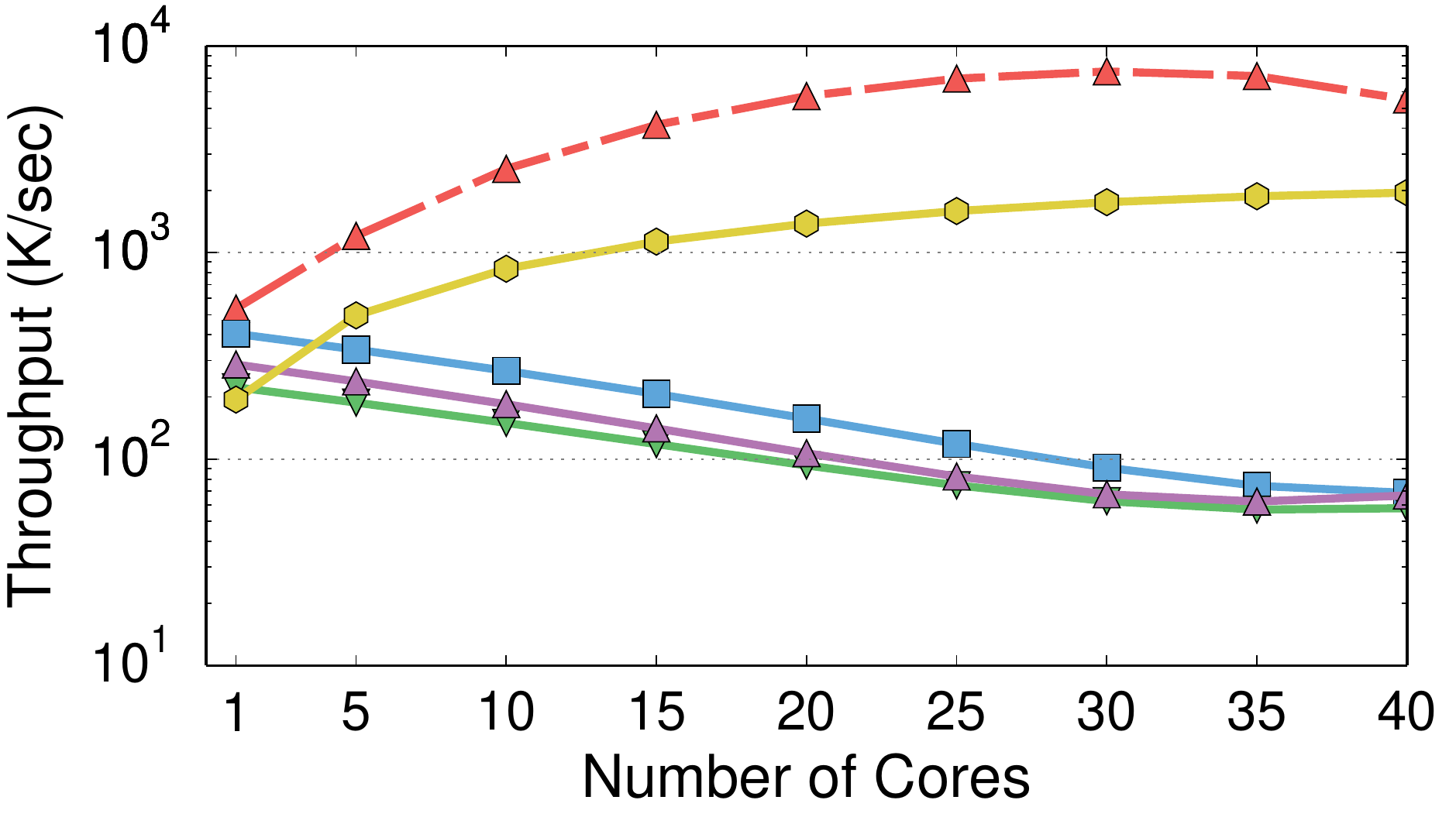}
    }       
    \caption{
       Throughput (K events per second) comparison of {different} applications under different schemes. 
    }
    \label{figures:use_case_results}
\end{figure}

The comparison results are shown in Figure~\ref{figures:use_case_results}, and there are three major observations.
First, 
\system significantly outperforms the second-best scheme in all applications at large core counts (i.e., 3.8 times over \pat for \textsl{GS}, 1.7 times over \pat for \textsl{SL}, 3.3 times over \pat for \textsl{OB}, and 4.8 times over \lal for \textsl{TP}). 
However, there is still a large room to improve \system to achieve the performance upper bound indicated by \textsl{No-Lock}.
Second, 
as expected, \system brings lower performance improvement when the workload has heavy data dependencies (e.g., \textsl{SL}). 
This is because it can only evaluate a subset of operation chains (whose dependencies are resolved) in parallel during each round. 
Third, 
\pat generally performs better than \lal and \lwm, as it avoids blocking when transactions access disjoint partitions.
However, \pat performs poorly for \textsl{TP} because the workload only has 100 unique keys, and transactions are still heavily contented in the same partition. Excessive access to partition locks causes further performance degradation making it perform even worse than \lal.
In contrast, \system is still able to exploit parallelism from a batch of transactions with a sufficiently large punctuation interval.

\subcompact
\subsection{Transaction Processing Time Breakdown}
\label{subsec:breakdown}
\noindent
\framebox{
    \parbox{\dimexpr\linewidth-4\fboxsep-4\fboxrule}{
        \textbf{Finding (2):}        
        The centralized lock permitting process results in serious contention. Our investigation reveals that prior schemes spend $\sim$80\% of their execution time on synchronization.
    }
}

\tony{
As discussed earlier in Figure~\ref{fig:motivate}, Section~\ref{sec:introduction}, state access overhead quickly dominates runtime.
We now use \textsl{SL} as a case to further study transaction processing time (including state access and access overhead) breakdown under different schemes.
}
Following the previous work~\cite{Yu:2014:SAE:2735508.2735511}, 
we report how much time is spent on different components in the processing of a state transaction.
1) \emph{Useful}: The time spent on accessing states.
2) \emph{Sync}: The time spent on synchronization.
It consists of blocking time before lock insertion is permitted in \lal, \lwm and \pat or blocking time due to synchronization barriers during mode switching in \system.
3) \emph{Lock}: The total amount of time that a transaction spends inserting locks after it is permitted to do so.
4) \emph{RMA}: The time spent on remote memory access. 
A thread may remotely access global counters in the case of \lal, \lwm, and \pat. 
\system may involve remote access during transaction decomposition as threads need to insert decomposed operation into appropriate operation chains.
Actual state access may also cause remote memory access for all schemes run on multi-sockets.
5) \emph{Others}: The time spent for all other operations and system overheads such as index lookup and context switching.

\begin{figure}[t]    
\centering
     \begin{center}
     \fbox{
     \includegraphics[width=0.7\columnwidth]
         {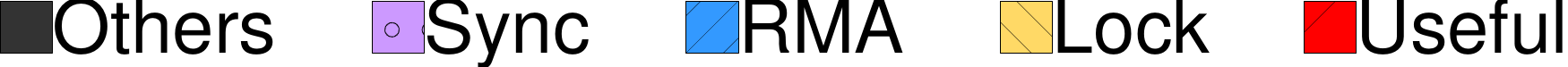}
     }
    \end{center}        
\captionsetup[subfigure]{font=scriptsize,labelfont=scriptsize}
    \subfloat[{Single socket (10 cores)}]{
        \includegraphics[width=0.38\textwidth]
            {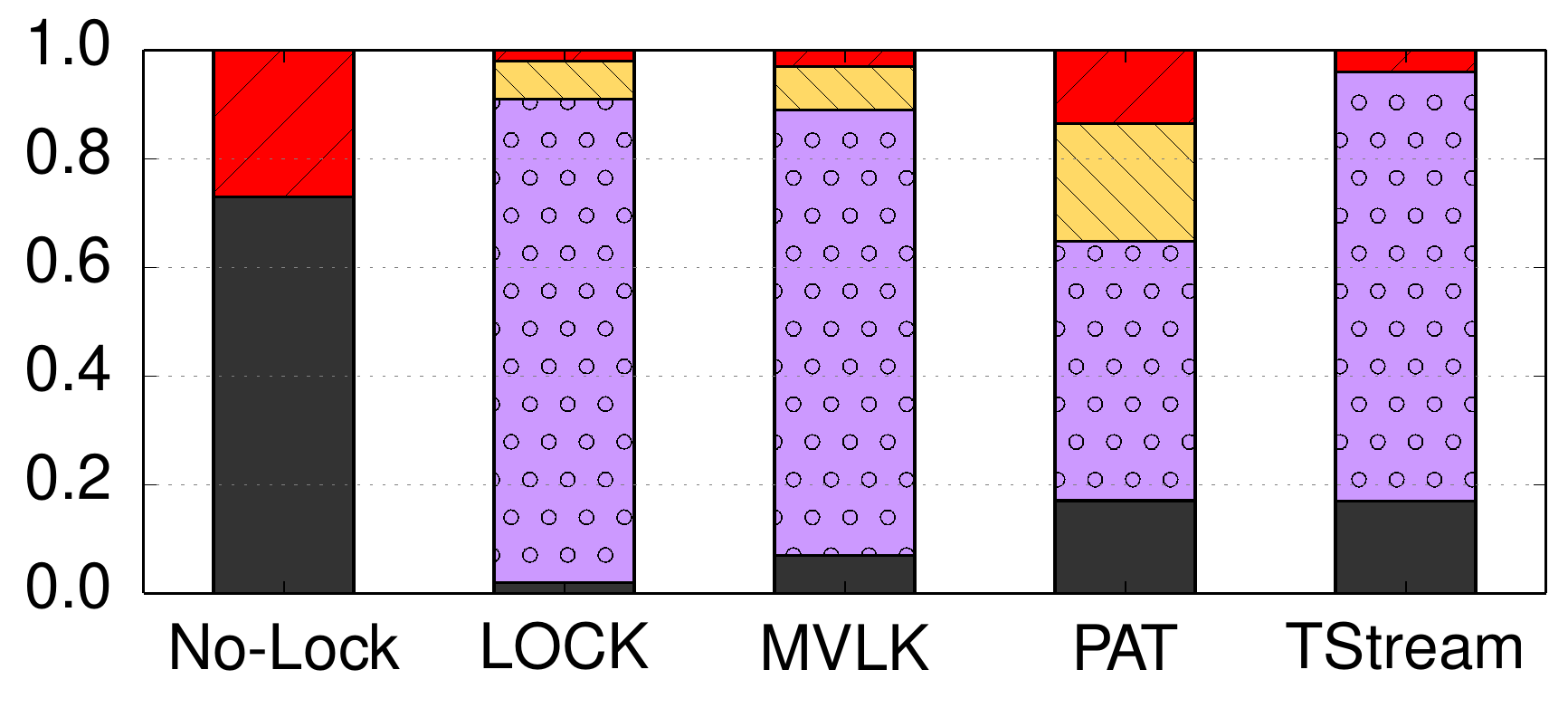}
    }    
    
    \subfloat[{Four sockets (40 cores)}]{
        \includegraphics[width=0.38\textwidth]
            {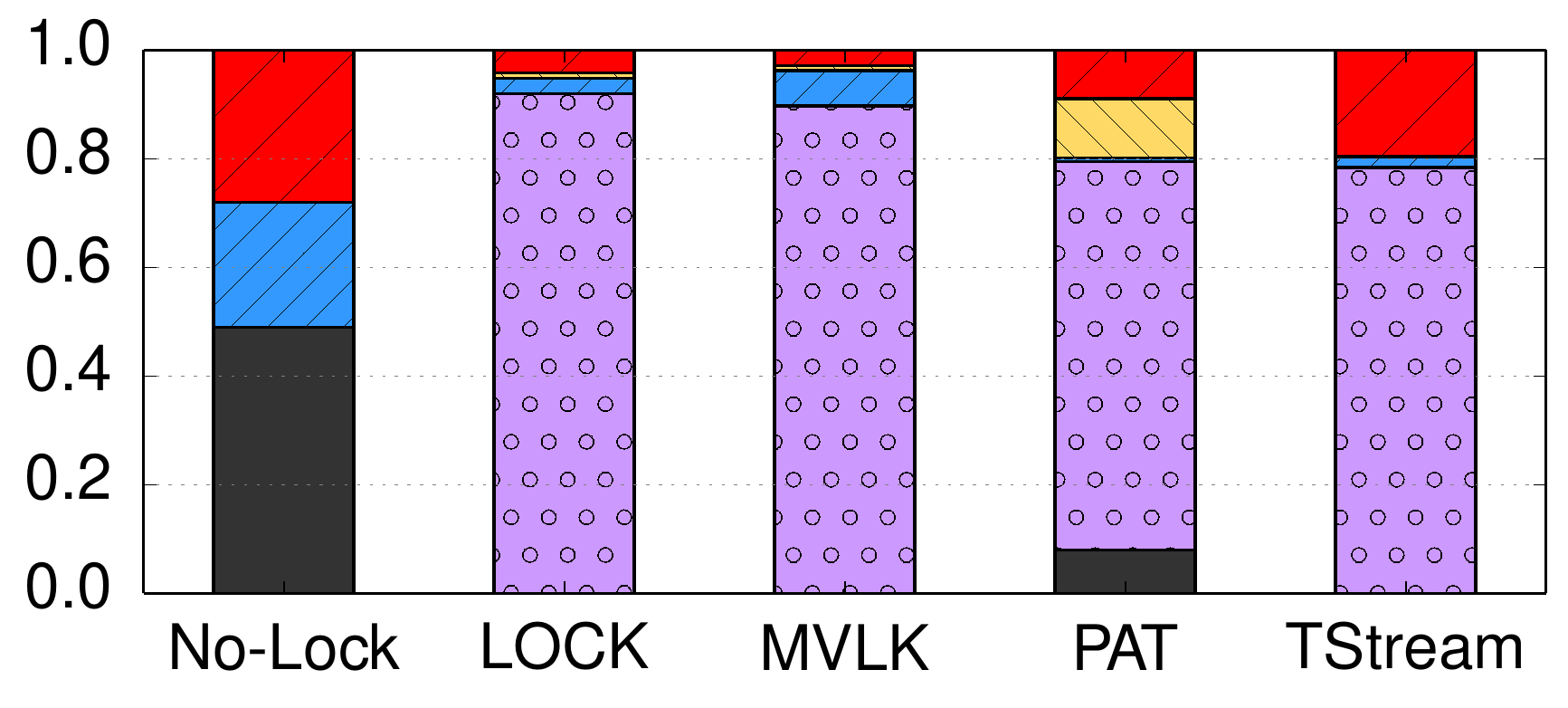}
    }         
    \caption{
       Runtime breakdown per state transaction in \textsl{SL}.
    }
    \label{figures:use_case_breakdown}
\end{figure}

Figure~\ref{figures:use_case_breakdown} shows the time breakdown when the system is run on a single or four CPU sockets. 
There are two major takeaways.
First, \textsl{No-Lock} spends more than 50\% of the time on \emph{Others}. 
Further investigation reveals that index lookup is the root cause of this performance degradation.
We defer the study of more scalable index design to future work and concentrate on concurrent execution control in this work.
Second, \emph{Sync} overhead dominates all consistency preserving schemes regardless of the effect of NUMA. 
Although \lwm spends less time in \emph{Sync} compared to \lal as read may not be blocked by write, 
it spends more time in reading and updating the \texttt{lwm} variables (grouped under the \emph{Others} overhead). 
\system shows a high synchronization overhead in \textsl{SL} due to heavy data dependencies. 
This shows that there is a large room for improvement.

\begin{figure}[t]       
\centering
     \begin{center}
     \fbox{
     \includegraphics[width=0.92\columnwidth]
         {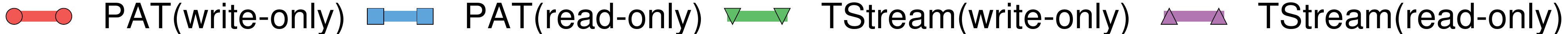}  
     }
    \end{center}  
    \centering
    \subfloat[Varying ratio of multi-partition txns (length=6).]{%
        \includegraphics*[width=0.38\textwidth]{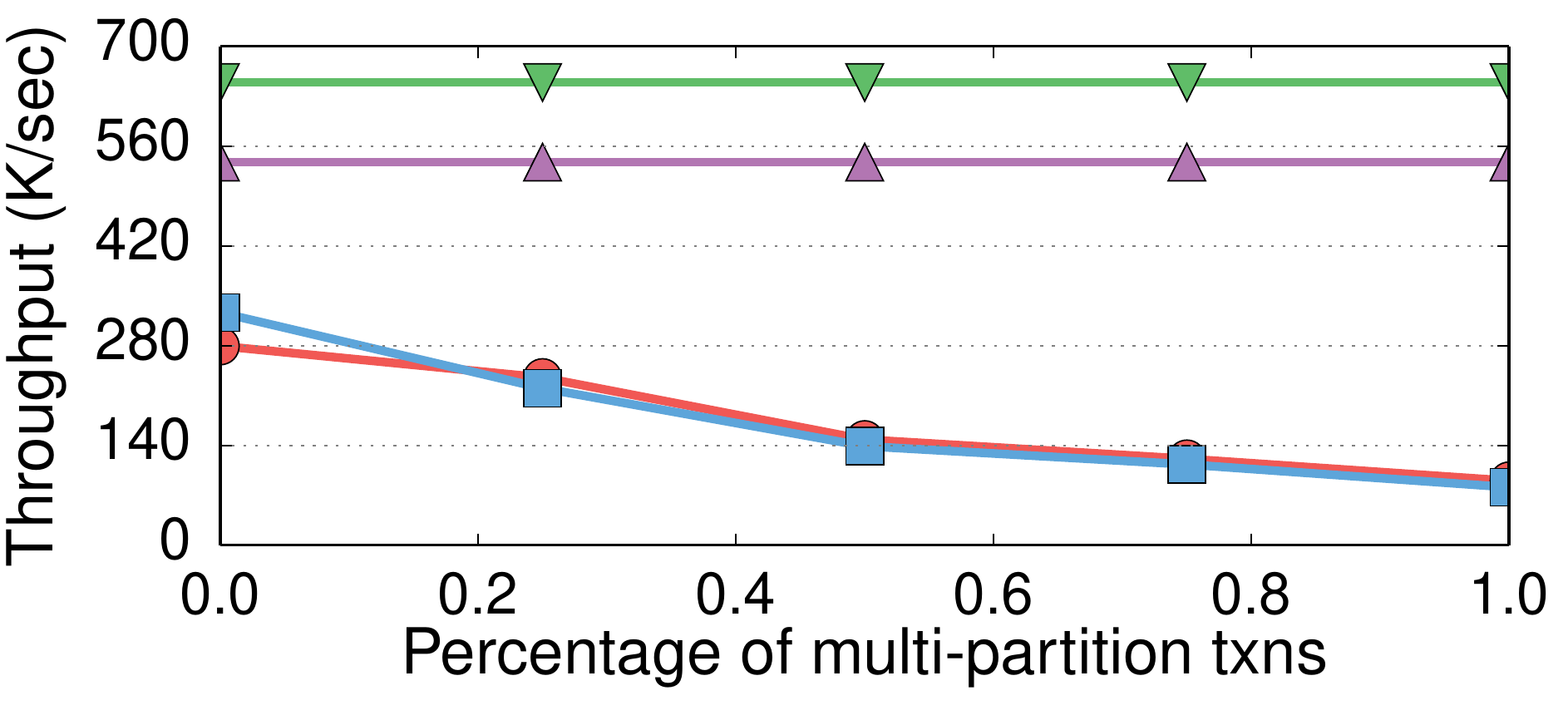}    \label{fig:multi_partition_r}
    }

    \subfloat[Varying length of multi-partition txns (ratio=50\%).]{%
         \includegraphics*[width=0.38\textwidth]{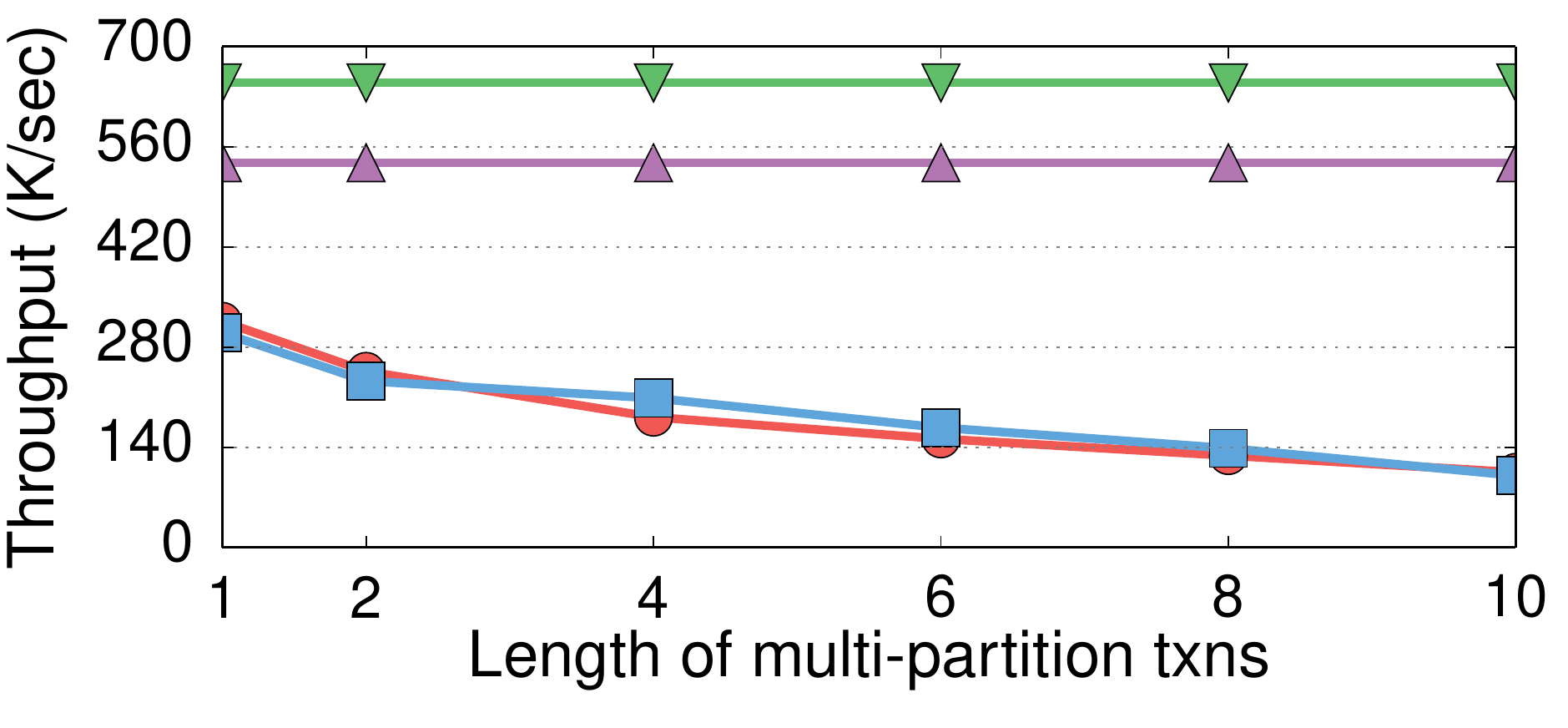}   \label{fig:multi_partition_n}
    }
    \caption{Multi-partition transaction evaluation.}
    \label{fig:multi_partition} 
\end{figure} 

\begin{figure}[t]
    \centering
     \begin{center}
     \fbox{
     \includegraphics[width=0.55\columnwidth]
         {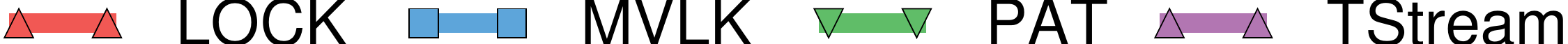}         
     }
    \end{center}   
    \subfloat[{Read/write workload ratio}.]{
            \includegraphics[width=0.38\textwidth]
                {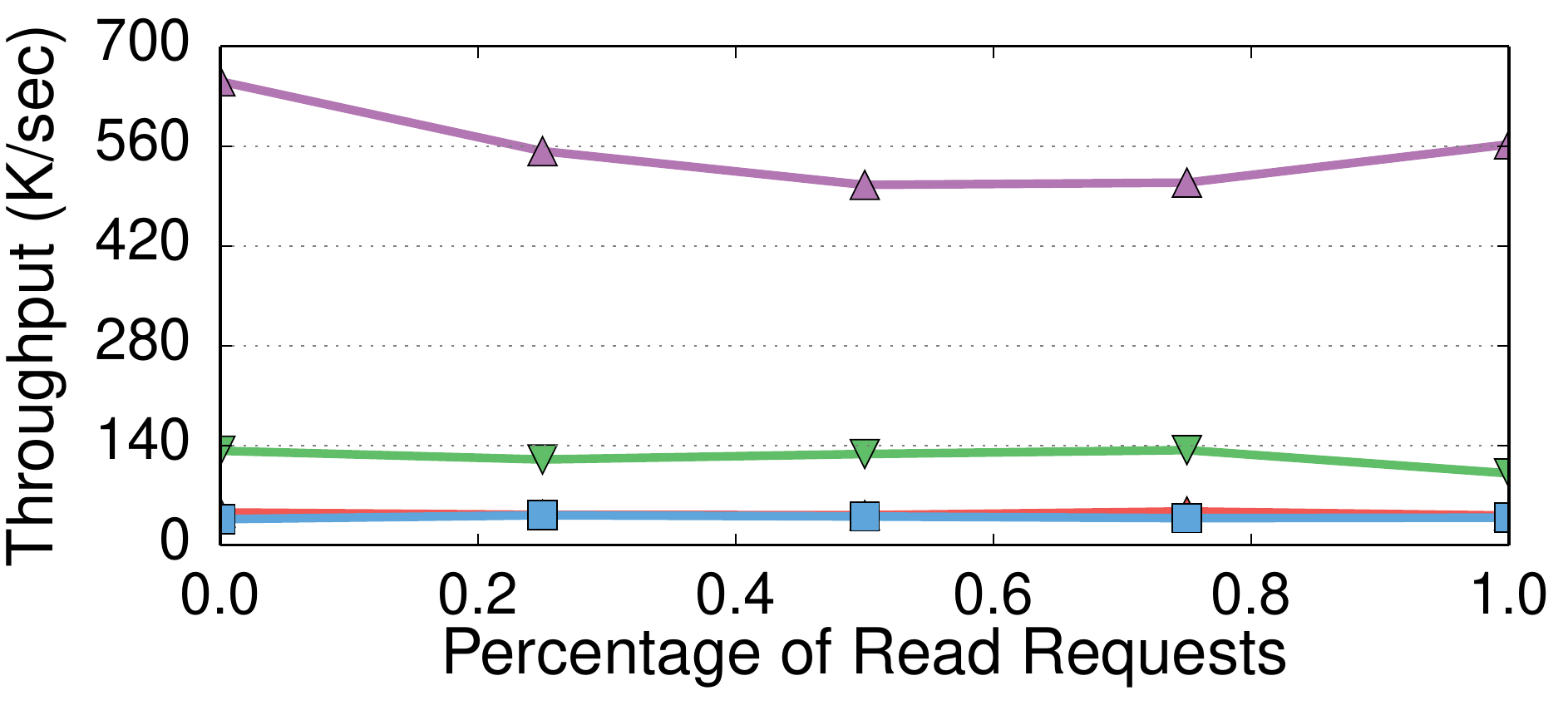}
         }   
            
    \subfloat[State access skewness.]{
        \includegraphics[width=0.38\textwidth]
            {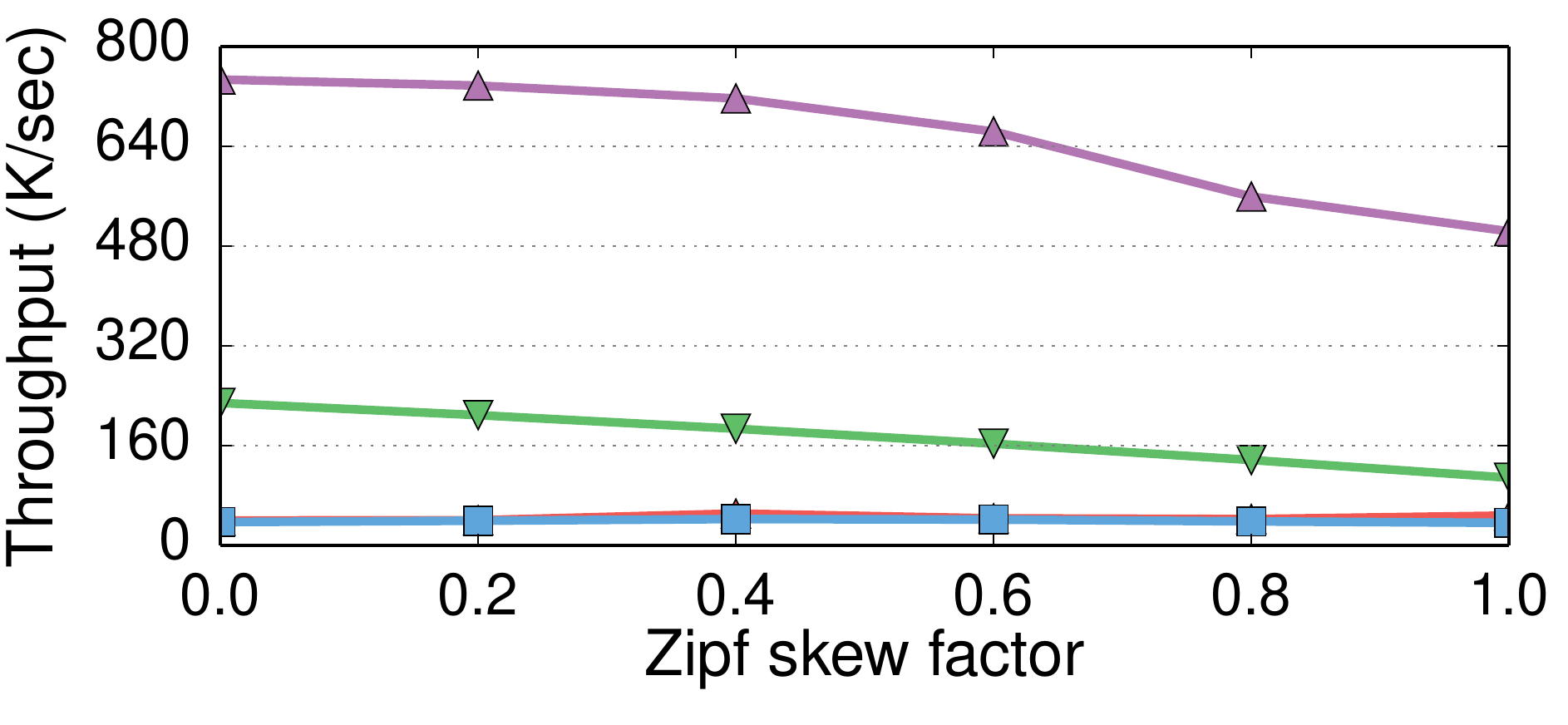}
    }               
    \caption{
       Varying application workload configurations of \textsl{GS}.
    }
    \label{figures:workload_senstivity}          
\end{figure}   
 
\begin{figure}[t]    
\centering
         \begin{center}
         \fbox{\parbox[t]{0.6\columnwidth}{
         \includegraphics[width=0.6\columnwidth]
             {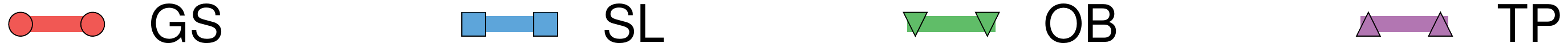}             
         }
         }
         \end{center}    
        \subfloat[Throughput]{%
             \includegraphics*[width=0.38\textwidth]{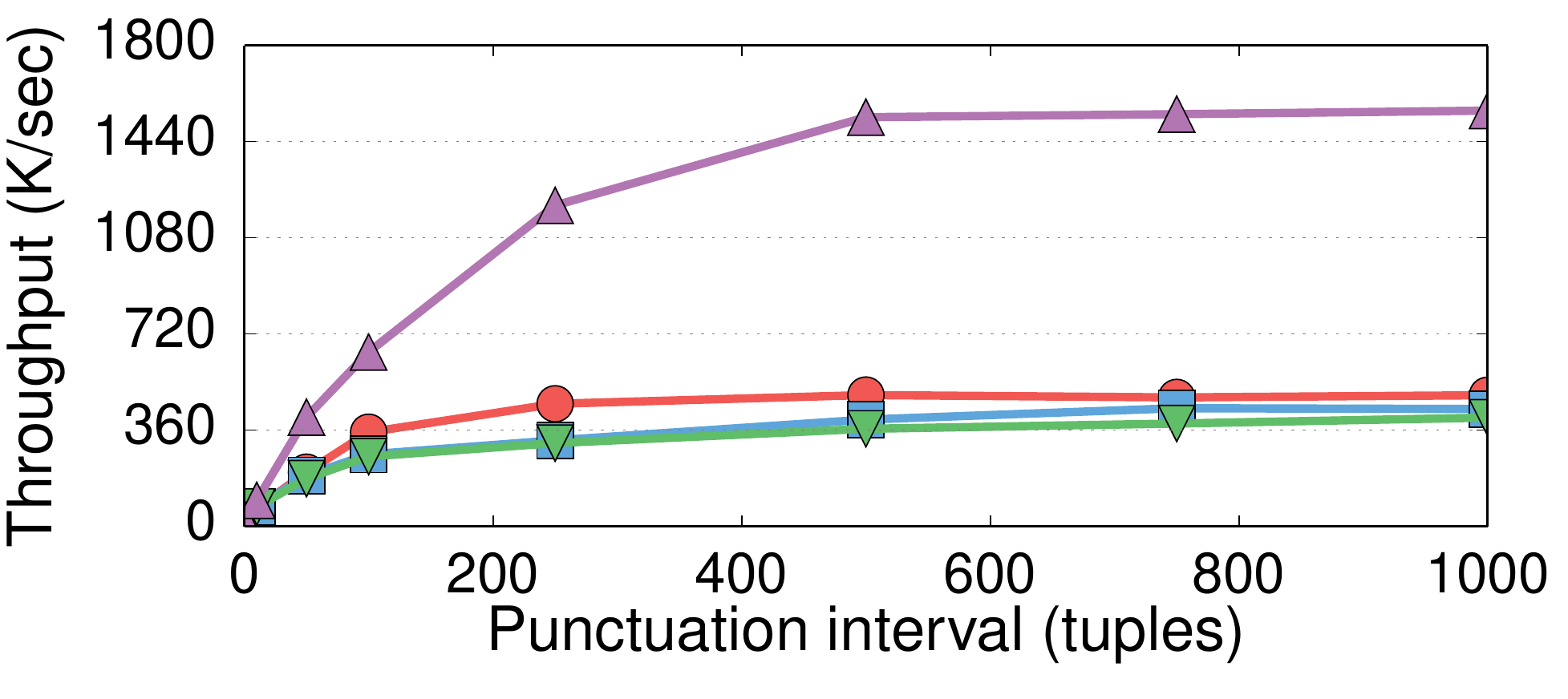}   
        }        
         
        \subfloat[$99^{th}$ percentile latency]{%
             \includegraphics*[width=0.38\textwidth]{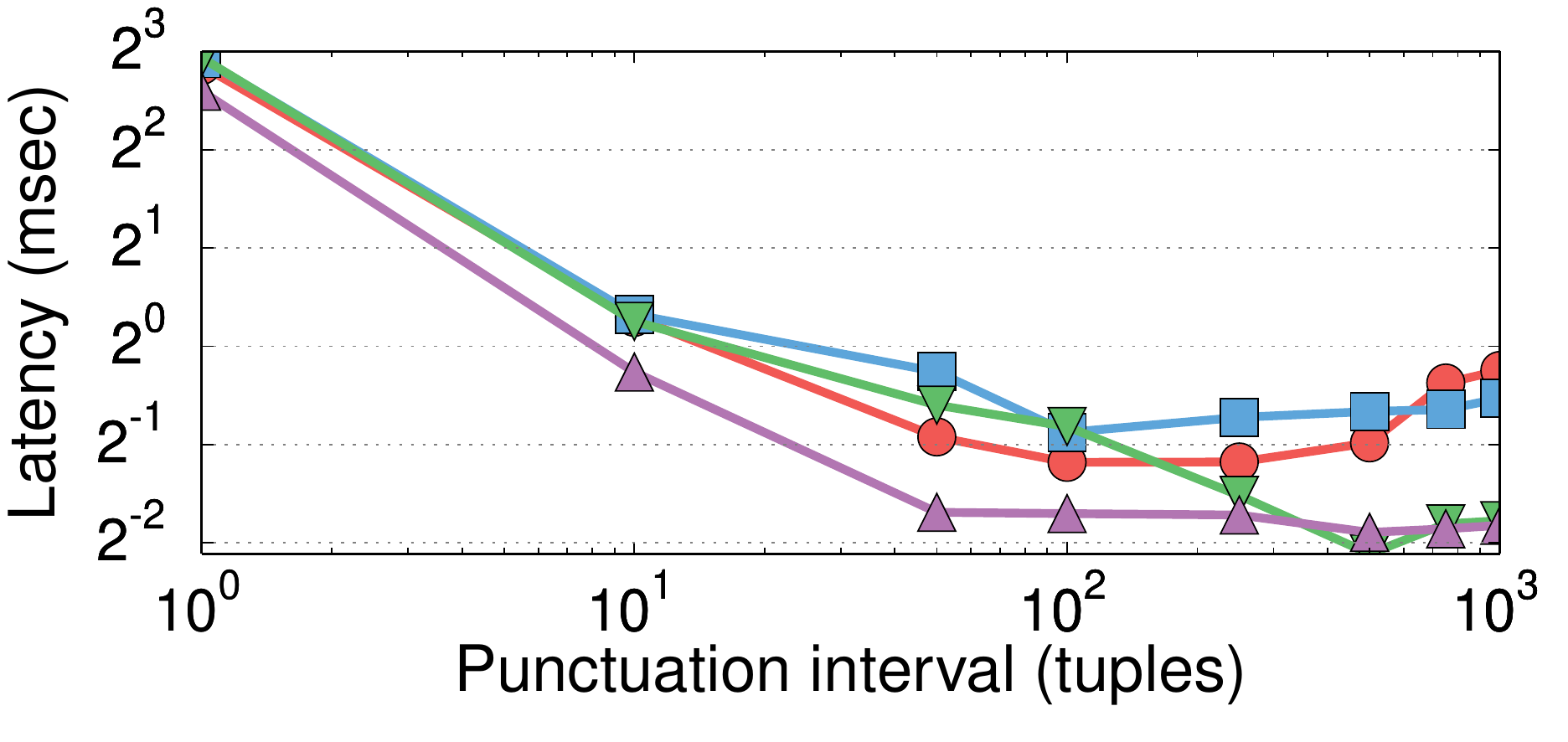}   
        }
        
    \caption{Effect of varying punctuation interval.}
    \label{fig:interval}             
\end{figure}

\subcompact
\subsection{Workload Sensitivity Study}
\label{subsec:single}
\noindent
\framebox{
    \parbox{\dimexpr\linewidth-4\fboxsep-4\fboxrule}{
        \textbf{Finding (3):}     
        The fine-grained design makes \system robust to different workloads. 
        Particularly, it maintains high performance under a) varying ratios and lengths of the multi-partition transaction, b) varying read/write state access ratios, and c) highly skewed access.
    }
}

We now use \textsl{GS} as an example to evaluate the different schemes under varying workload configurations.

\textbf{Multi-partition Transaction Percentage.}
\label{subsec:multi}
We first study the effect of state partitioning. 
We use a simple hashing strategy to assign states to partitions based on their primary keys so that each partition stores a similar number of states. 
As a common issue of all partition-based algorithms~\cite{Pavlo2012}, 
the performance of \pat is heavily dependent on the length and ratio of multi-partition transactions. 
We first configure each multi-partition transaction to access six different partitions of the application states. 
We then vary the percentage of multi-partition transactions in the workload. The results are shown in Figure~\ref{fig:multi_partition} (a). 
There are two key observations.
First, since \pat is specially designed to take advantage of partitioning, it has low synchronization overhead when no multi-partition transactions are present (i.e., ratio=0\%). 
However, it performs worse than \system even without any multi-partition transaction as \system can utilize more parallelism opportunities due to its fine-grained execution paradigm. 
Second, \pat's performance degrades with more multi-partition transactions as it further reduces parallelism opportunities.
A similar observation can be found in Figure~\ref{fig:multi_partition} (b), where we vary the length of multi-partition transactions and fix its ratio to 50\%. 
In the following studies, we set the multi-partition ratio to 50\% under \pat. 

\textbf{Read Request Percentage.}
We now vary the percentage of events that trigger read requests to application states from 0\% (write-only) to 100\% (read-only).
In this study, 
we remove the summation computation from \textsl{GS} and focus on evaluating the efficiency of state access.
We also set the key skew factor to be 0, and hence   states are accessed with uniform frequency.
Figure~\ref{figures:workload_senstivity} (a) shows the results and there are two major observations.
First, varying read/write request ratio has a minor effect on system performance under prior schemes, \lal, \lwm and \pat. 
This is because their execution runtime is dominated by synchronization overhead. 
Second, \system generally performs worse with more read requests as \system has to write the state value to \emph{EventBlotter} of the triggering event (which triggers the read request) during transaction evaluation. 
An interesting point to take note is that \system's performance increases slightly under the read-only workload compared to the mixed workload. 
When there are both reads and writes to the same state, hardware prefetchers are not effective as each prefetch can steal read and write permissions for shared blocks from other processors, leading to permission thrashing and overall performance degradation~\cite{Prefetching2}.


\textbf{State Access Skewness.}
In this study, we configure a write-only workload to examine how different schemes perform under contented state updates.
Figure~\ref{figures:workload_senstivity}(b) shows that \system is tolerant to access skewness. 
Prior schemes perform worse with increasing skewness as there is more intensive contention on the same lock.
In contrast, \system achieves high performance even under serious skewness because \system is still able to discover parallelism opportunities among a batch of transactions (a punctuation interval of 500).


\begin{figure}[t]
\centering  
    \begin{minipage}{0.38\textwidth}        
    \includegraphics*[width=\textwidth]{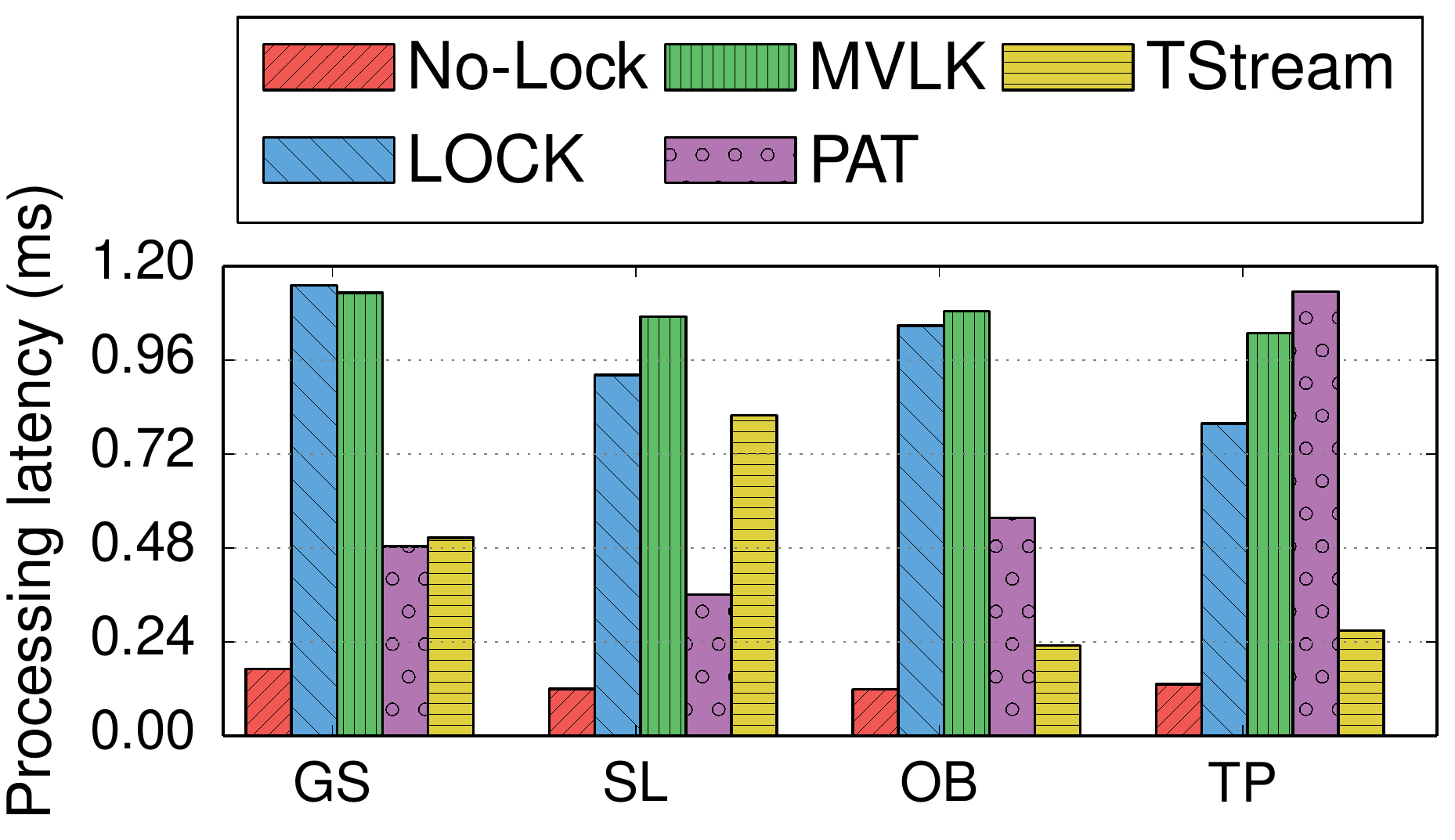} 
    \caption{$99^{th}$ percentile end-to-end processing latency.}   
    \label{fig:latency_compare}
    \end{minipage}        
\end{figure}

\begin{figure}[t]
\centering 
    \begin{minipage}{0.38\textwidth}        
    \includegraphics*[width=\textwidth]{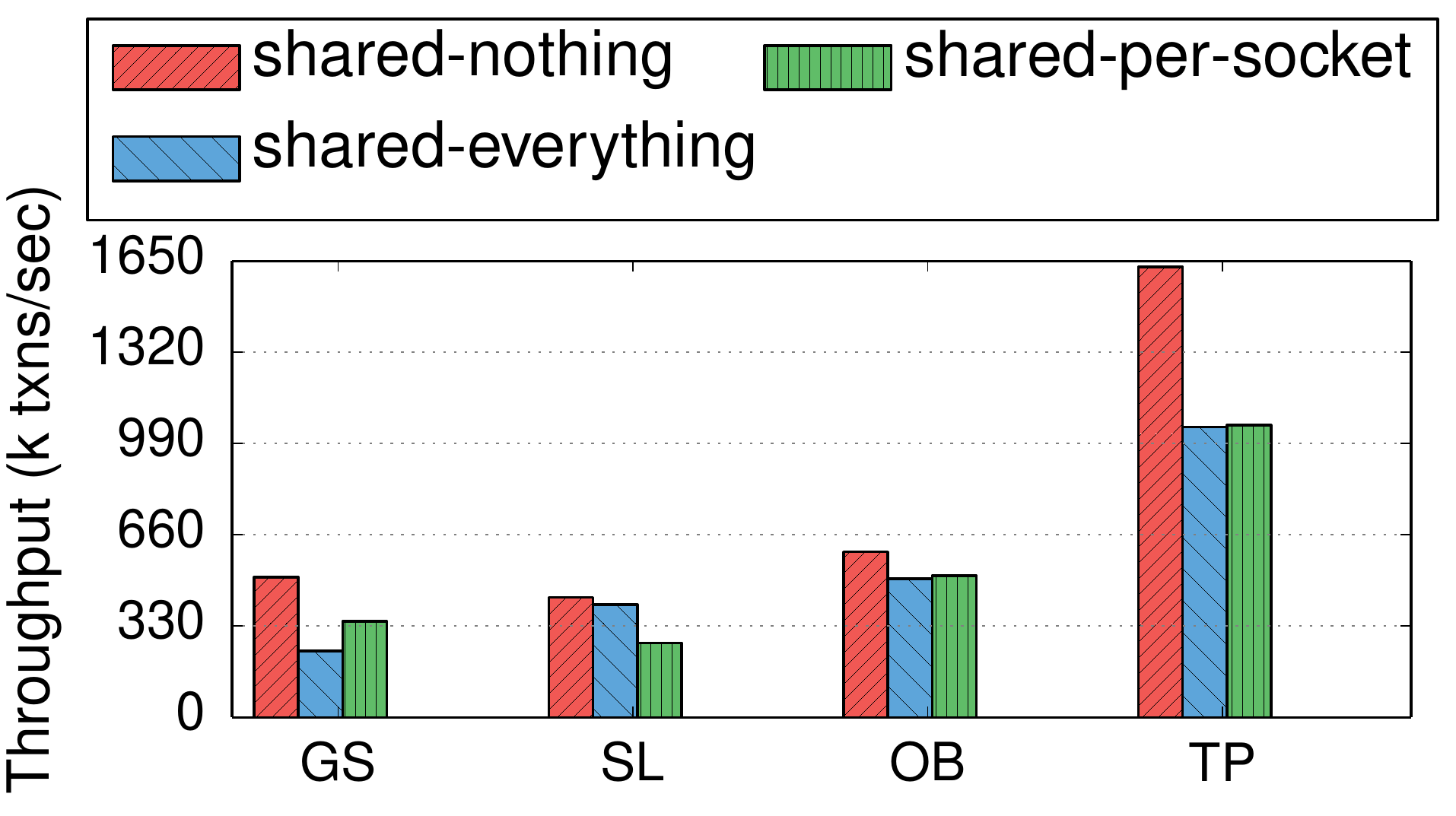} 
    \caption{Varying NUMA-aware configurations.}   
    \label{fig:numaaware}
    \end{minipage}      
\end{figure}

\subcompact
\subsection{System Sensitivity Study}
\label{subsec:factor}
\noindent
\framebox{
    \parbox{\dimexpr\linewidth-4\fboxsep-4\fboxrule}{
        \textbf{Finding (4):}     
        \tony{\system can be tuned to achieve low processing latency and high throughput.}
        We also find that the shared-nothing NUMA-aware configuration achieves the best performance.
    }
}

 
\textbf{Varying Punctuation Interval.}
The number of transactions to handle between two consecutive punctuation plays a critical role in \system's performance. 
Figure~\ref{fig:interval} (a) shows that the performance of \system generally increases with a larger punctuation interval. 
It also shows that a large punctuation interval is especially beneficial for
\textsl{TP}  because the workload has only 100 unique segment IDs and transactions are heavily contented at the same state. 
\tony{By allowing more transactions to be accumulated, \system increases parallelism opportunities among more decomposed operations, and  its performance hence increases significantly.}
Figure~\ref{fig:interval} (b) shows the processing latency of \system with various punctuation intervals. 
Following the previous work~\cite{adaptivebatch}, we define the end-to-end processing latency as the duration between the time when an input event enters the system and the time when the result is generated. 
Thanks to the significantly improved performance, \system achieves very low processing latency.
When the punctuation interval is set to 500, its 99$^{th}$-percentile processing latency is around 0.23$\sim$0.63 ms, which satisfies many existing use cases~\cite{adaptivebatch}. 
It also shows that there is no clear trade-off between throughput and latency under varying punctuation intervals. This is because higher throughput also reduces queuing delays. Latency increases with increasing punctuation interval only when throughput can not be further improved (e.g., \textsl{GS} at an interval of  250).
\tony{
Figure~\ref{fig:latency_compare} further shows that \system (punctuation interval=500) achieves comparable and sometimes even lower processing latency compared to the state-of-the-art. 
The optimal (e.g., maximum throughput) punctuation interval may be affected by many factors including machine characteristics (e.g., number of cores, size of LLC, and memory), number of unique keys in the workload, state size, tuple size, length of the state transaction, etc. Due to its considerable complexity, we leave the estimation of the optimal punctuation interval itself to future work. 
}

\textbf{Effect of NUMA-aware Optimizations.}
We now compare different NUMA-aware processing configurations of \system including shared-nothing, shared-everything, and shared-per-socket. 
Work-stealing can be further enabled in the latter two configurations and our experimental results show that work-stealing significantly improves their throughput by 1.6$\sim$7.0 times.
However, Figure~\ref{fig:numaaware} shows that \system achieves the best performance for all applications under the shared-nothing configuration.  
This indicates that cross-core and cross-socket communication during state transaction processing should always be  avoided. 
Nevertheless, we plan to investigate this impact on other applications that may be more sensitive to workload imbalance rather than communication overhead.

\subsection{Compared to SStore}
\label{subsec:sanity}
For a sanity test, we have compared SStore's performance against TStream (\pat scheme) on single core using SStore's Micro-Benchmark with one stored procedure setting (Fig.6 ~\cite{S-Store}), which contains three write operations. 
The results show that \system (\pat scheme) outperforms SStore about three times: 
SStore achieves a throughput of $\sim$ 3.6K events/sec and \system (\pat scheme) achieves $\sim$ 11.7K events/sec. 
This validates the efficiency of our re-implementation.
The performance superiority comes from \system's more efficient execution mechanism. 
For example, three write operations in this test are consecutively executed by one thread in \system rather than passively triggered by triggers (SStore) and context switching overhead is significantly reduced.

%% file: related.tex
\section{Related Work}
\label{sec:related}
%

\textbf{Concurrent Stateful Stream Processing.}
We have reviewed some of the related work in Section~\ref{sec:background} and now discuss a few more.
Botan et al.~\cite{botan2012transactional} presented an \textit{unified transactional model} for streaming applications.
Affetti et al.~\cite{Affetti:2017:FIS:3093742.3093929} recently proposed a state consistency model for stream processing. 
Both studies provide the same formal definitions on how mutable application states can be shared among executors during stream processing through transactional semantics, and we have adopted their consistency model.
However, their implementations heavily rely on locks to guarantee state consistency.
Unless these systems, \system's novel design has been shown to achieve much higher throughput and scalability with various workloads. 
There is also a recent commercial system, called Streaming Ledger~\cite{Transactions2018} for extending Flink to support concurrent state access with a  goal similar to ours.
It is close-sourced, and we can not compare our system with that. 

{\textbf{Database Partitioning.}}
Prior work~\cite{Thomson:2010:CDD:1920841.1920855, hyper} propose to divide the underlying storage into multiple logical partitions, each of which is assigned a single-thread execution engine with exclusive access. 
Transaction workloads in those databases are partitioned according to the primary key(s) in the root table~\cite{Stonebraker:2007:EAE:1325851.1325981}, and the performance can significantly degrade as the ratio of multi-partition transactions increases~\cite{Porobic:2012:OHI:2350229.2350260}. 
S-Store~\cite{S-Store} adopts the same technique with extensions to further guarantee state access ordering~\cite{S-Store-demo}.
The partition-based approach's common drawback is their handling of multi-partition transactions.
In contrast, \system decomposes a collection of transactions at runtime and execute the resulting operation chains at  high system concurrency. 

\textbf{Program Partitioning.} 
Many have proposed adopting program partitioning and transformation to optimize 
the performance of transaction processing,
such as~\cite{Bernstein:1999:CCS:337919.337922}. 
TStream deviates from existing techniques such as transaction chopping~\cite{Shasha:1995:TCA:211414.211427} which are purely static.
\system dynamically restructures potentially conflicting operations in a collection of state transactions into independent groups called operation chains which are evaluated in a determined sequence (\textbf{F3}).
Transaction-chopping and its many variants such as~\cite{Narula:2014:PRC:2685048.2685088} were proposed in the context of nondeterministic transaction processing, and thus their program partitioning technique does not account for the state access sequence that is necessary for state consistency of stream processing.
The periodic transaction processing of \system is in the spirit of lazy transaction evaluation~\cite{Faleiro:2014:LET:2588555.2610529}, 
but \system needs to ensure that transactions are processed following input event sequence,
which results in different optimization opportunities (e.g., sorted operation chains).

\textbf{Multicore Architectures.}
Multicore architectures have brought many research challenges and opportunities for in-memory data management, as outlined in recent surveys~\cite{inmemorybigdata}. 
To meet the increasing performance demand, optimizing stream processing on multicore machines has been a hot research topic~\cite{profile,Zeuch:2019:AES:3303753.3316441,SABER}.
\system is built to improve multicore utilization standing on the shoulders of many valuable existing works such as~\cite{briskstream,StreamBox,Porobic:2012:OHI:2350229.2350260}.  
However, none of the previous work addresses the scalability bottlenecks that \system solves, 
i.e. how to scale concurrent state access in stream processing with consistency guarantee.

%% file: conclusion.tex
\section{Conclusion}
\label{sec:conclude}
With the increasing adoption of stream processing in emerging use cases, we believe that an efficient concurrent stateful DSPS becomes more and more desirable.
\system demonstrates that efficient concurrent state access during stream processing can be elegantly supported with its novel dual-mode scheduling and dynamic restructuring execution mechanism on modern multicore architectures. 
In particular, it guarantees strict state consistency, while judiciously exploits more parallelism opportunities -- both within the processing of each input event and among a (tunable) batch of input events. 

%% file: report.bbl
\begin{thebibliography}{10}
\providecommand{\url}[1]{#1}
\csname url@samestyle\endcsname
\providecommand{\newblock}{\relax}
\providecommand{\bibinfo}[2]{#2}
\providecommand{\BIBentrySTDinterwordspacing}{\spaceskip=0pt\relax}
\providecommand{\BIBentryALTinterwordstretchfactor}{4}
\providecommand{\BIBentryALTinterwordspacing}{\spaceskip=\fontdimen2\font plus
\BIBentryALTinterwordstretchfactor\fontdimen3\font minus
  \fontdimen4\font\relax}
\providecommand{\BIBforeignlanguage}[2]{{%
\expandafter\ifx\csname l@#1\endcsname\relax
\typeout{** WARNING: IEEEtran.bst: No hyphenation pattern has been}%
\typeout{** loaded for the language `#1'. Using the pattern for}%
\typeout{** the default language instead.}%
\else
\language=\csname l@#1\endcsname
\fi
#2}}
\providecommand{\BIBdecl}{\relax}
\BIBdecl

\bibitem{flink}
(2018) Apache flink, \url{https://flink.apache.org/}.

\bibitem{Storm}
(2018) Apache storm, \url{http://storm.apache.org/}.

\bibitem{heron}
S.~Kulkarni and et~al., ``Twitter heron: Stream processing at scale,'' in
  \emph{SIGMOD '15}.

\bibitem{DBLP:conf/icde/WuT15}
Y.~Wu and K.~Tan, ``Chronostream: Elastic stateful stream computation in the
  cloud,'' in \emph{ICDE'15}.

\bibitem{Carbone:2017:SMA:3137765.3137777}
P.~Carbone and et~al., ``State management in apache flink: Consistent stateful
  distributed stream processing,'' \emph{Proc. VLDB Endow. 2017}.

\bibitem{partition_cost}
N.~R. Katsipoulakis and et~al., ``A holistic view of stream partitioning
  costs,'' \emph{Proc. VLDB Endow. 2017}.

\bibitem{acep}
D.~Wang and et~al., ``Active complex event processing over event streams,''
  \emph{Proc. VLDB Endow. 2011}.

\bibitem{Affetti:2017:FIS:3093742.3093929}
L.~Affetti and et~al., ``Flowdb: Integrating stream processing and consistent
  state management,'' in \emph{DEBS '17}.

\bibitem{botan2012transactional}
I.~Botan and et~al., ``Transactional stream processing,'' in \emph{EDBT '12}.

\bibitem{S-Store}
J.~Meehan and et~al., ``S-store: Streaming meets transaction processing,''
  \emph{Proc. VLDB Endow. 2015}.

\bibitem{Linear}
A.~Arasu and et~al., ``Linear road: A stream data management benchmark,'' in
  \emph{VLDB '04}.

\bibitem{briskstream}
S.~Zhang and et~al., ``Briskstream: Scaling data stream processing on
  shared-memory multicore architectures,'' in \emph{SIGMOD '19}.

\bibitem{Aurora}
D.~J. Abadi and et~al., ``Aurora: A new model and architecture for data stream
  management,'' \emph{The VLDB Journal, 2003}.

\bibitem{profile}
S.~Zhang and et~al., ``Revisiting the design of data stream processing systems
  on multi-core processors,'' in \emph{ICDE'17}.

\bibitem{Aeolus}
M.~J. Sax and M.~Castellanos, ``Building a transparent batching layer for
  storm.''\hskip 1em plus 0.5em minus 0.4em\relax HP Labs Technical Report,
  2013.

\bibitem{Pavlo2012}
A.~Pavlo and et~al., ``Skew-aware automatic database partitioning in
  shared-nothing, parallel oltp systems,'' in \emph{SIGMOD '12}.

\bibitem{bernstein2009principles}
P.~A. Bernstein and E.~Newcomer, \emph{Principles of transaction
  processing}.\hskip 1em plus 0.5em minus 0.4em\relax Morgan Kaufmann, 2009.

\bibitem{Yu:2016:TTT:2882903.2882935}
X.~Yu and et~al., ``Tictoc: Time traveling optimistic concurrency control,'' in
  \emph{SIGMOD '16}.

\bibitem{Yu:2014:SAE:2735508.2735511}
X.~Yu and et~al., ``Staring into the abyss: An evaluation of concurrency
  control with one thousand cores,'' \emph{Proc. VLDB Endow. 2014}.

\bibitem{Wu2016}
Y.~Wu and et~al., ``Transaction healing: Scaling optimistic concurrency control
  on multicores,'' in \emph{SIGMOD '16}.

\bibitem{he2008mars}
B.~He and et~al., ``Mars: a mapreduce framework on graphics processors,'' in
  \emph{PACT'08}.

\bibitem{Tucker:2003:EPS:776752.776780}
P.~A. Tucker and et~al., ``Exploiting punctuation semantics in continuous data
  streams,'' \emph{TKDE'03}.

\bibitem{StreamBox}
H.~Miao and et~al., ``Streambox: Modern stream processing on a multicore
  machine,'' in \emph{{USENIX} {ATC}'17}.

\bibitem{trill}
B.~Chandramouli and et~al., ``Trill: A high-performance incremental query
  processor for diverse analytics,'' \emph{Proc. VLDB Endow. 2014}.

\bibitem{barrier}
{Cyclicbarrier}.
  \url{https://docs.oracle.com/javase/7/docs/api/java/util/concurrent/CyclicBarrier.html}.

\bibitem{pugh1989skip}
W.~Pugh, ``Skip lists: A probabilistic alternative to balanced trees,'' in
  \emph{Workshop on Algorithms and Data Structures, 1989}.

\bibitem{Porobic:2012:OHI:2350229.2350260}
D.~Porobic and et~al., ``Oltp on hardware islands,'' \emph{Proc. of the VLDB
  Endow. 2012}.

\bibitem{6816692}
D.~{Porobic} and et~al., ``Atrapos: Adaptive transaction processing on hardware
  islands,'' in \emph{ICDE'14}.

\bibitem{Blumofe:1999:SMC:324133.324234}
R.~D. Blumofe and C.~E. Leiserson, ``Scheduling multithreaded computations by
  work stealing,'' \emph{J. ACM, 1999}.

\bibitem{Hirzel:2014:CSP:2597757.2528412}
M.~Hirzel and et~al., ``A catalog of stream processing optimizations,''
  \emph{ACM Comput. Surv. 2014}.

\bibitem{Zeuch:2019:AES:3303753.3316441}
S.~Zeuch and et~al., ``Analyzing efficient stream processing on modern
  hardware,'' \emph{Proc. VLDB Endow. 2019}.

\bibitem{Gray:1992:BHD:530588}
J.~Gray, \emph{Benchmark Handbook: For Database and Transaction Processing
  Systems}.\hskip 1em plus 0.5em minus 0.4em\relax San Francisco, CA, USA:
  Morgan Kaufmann Publishers Inc., 1992.

\bibitem{Transactions2018}
``{Data Artisans Streaming Ledger Serializable ACID Transactions on Streaming
  Data, \url{https://www.da-platform.com/streaming-ledger}},'' 2018.

\bibitem{bidding}
J.~Tan and M.~Zhong, ``An online bidding system (obs) under price match
  mechanism for commercial procurement,'' \emph{Applied Mechanics and
  Materials, 2014}.

\bibitem{mlc}
(2018) Intel memory latency checker, \url{https://software.intel.com/
  articles/intelr-memory-latency-checker}.

\bibitem{Prefetching2}
N.~D.~E. {Jerger} and et~al., ``Friendly fire: understanding the effects of
  multiprocessor prefetches,'' in \emph{ISPASS'06}.

\bibitem{adaptivebatch}
T.~Das and et~al., ``Adaptive stream processing using dynamic batch sizing,''
  in \emph{SOCC '14}.

\bibitem{Thomson:2010:CDD:1920841.1920855}
A.~Thomson and D.~J. Abadi, ``The case for determinism in database systems,''
  \emph{Proc. VLDB Endow. 2010}.

\bibitem{hyper}
A.~{Kemper} and T.~{Neumann}, ``Hyper: A hybrid oltp olap main memory database
  system based on virtual memory snapshots,'' in \emph{ICDE'11}.

\bibitem{Stonebraker:2007:EAE:1325851.1325981}
M.~Stonebraker and et~al., ``The end of an architectural era: (it's time for a
  complete rewrite),'' in \emph{VLDB '07}.

\bibitem{S-Store-demo}
U.~Cetintemel and et~al., ``S-store: A streaming newsql system for big velocity
  applications,'' \emph{Proc. VLDB Endow. 2014}.

\bibitem{Bernstein:1999:CCS:337919.337922}
A.~J. Bernstein and et~al., ``Concurrency control for step-decomposed
  transactions,'' \emph{Inf. Syst. 1999}.

\bibitem{Shasha:1995:TCA:211414.211427}
D.~Shasha and et~al., ``Transaction chopping: Algorithms and performance
  studies,'' \emph{ACM Trans. Database Syst. 1995}.

\bibitem{Narula:2014:PRC:2685048.2685088}
N.~Narula and et~al., ``Phase reconciliation for contended in-memory
  transactions,'' in \emph{OSDI'14}.

\bibitem{Faleiro:2014:LET:2588555.2610529}
J.~M. Faleiro and et~al., ``Lazy evaluation of transactions in database
  systems,'' in \emph{SIGMOD '14}.

\bibitem{inmemorybigdata}
H.~Zhang and et~al., ``In-memory big data management and processing: A
  survey,'' \emph{TKDE'15}.

\bibitem{SABER}
A.~Koliousis and et~al., ``Saber: Window-based hybrid stream processing for
  heterogeneous architectures,'' in \emph{SIGMOD '16}.

\end{thebibliography}
